\documentclass[twocolumn]{aastex63}

\usepackage{graphicx}
\usepackage[toc,page]{appendix}
\usepackage{slashed}
\usepackage{verbatim}
\usepackage[caption = false]{subfig}
\usepackage{amsmath}                           
\usepackage{amssymb}

\newcommand{\ie}{{\it i.e.}}

\newcommand{\eg}{{\it e.g.}}

\newcommand{\cf}{{\it cf.}}

\newcommand{\eq}{Eq.}

\newcommand{\Sec}{Section}

\newcommand{\App}{Appendix}

\newcommand{\Tab}{Tab.}

\newcommand{\equ}[1]{\eq~(\ref{equ:#1})}
\newcommand{\sect}[1]{\Sec~(\ref{sec:#1})}
\newcommand{\appr}[1]{\App~(\ref{app:#1})}
\newcommand{\figu}[1]{Fig.~\ref{fig:#1}}
\newcommand{\tabl}[1]{\Tab~\ref{tab:#1}}

\definecolor{violet}{rgb}{0.858, 0.188, 0.478}

\newcommand{\revise}[1]{{#1}}

\graphicspath{{Plots/}}

\begin{document}

\title{Impact of the collision model on the multi-messenger emission
from Gamma-Ray Burst internal shocks}

\correspondingauthor{Annika Rudolph}
\email{annika.rudolph@desy.de}

\author{Annika Rudolph}
\affiliation{Deutsches Elektronen Synchrotron (DESY), Platanenallee 6, D-15738 Zeuthen, Germany}

\author{Jonas Heinze}
\affiliation{Deutsches Elektronen Synchrotron (DESY), Platanenallee 6, D-15738 Zeuthen, Germany}

\author{Anatoli Fedynitch}
\affiliation{Deutsches Elektronen Synchrotron (DESY), Platanenallee 6, D-15738 Zeuthen, Germany}
\affiliation{Dept.\ of Physics, University of Alberta, Edmonton, Alberta, Canada T6G 2E1}
\affiliation{Institute for Cosmic Ray Research, the University of Tokyo,
5-1-5 Kashiwa-no-ha, Kashiwa, Chiba 277-8582, Japan}

\author{Walter Winter}
\affiliation{Deutsches Elektronen Synchrotron (DESY), Platanenallee 6, D-15738 Zeuthen, Germany}

\begin{abstract}
	\vspace*{0.5cm}
	We discuss the production of multiple astrophysical messengers (neutrinos, cosmic rays, gamma-rays) in the Gamma-Ray Burst (GRB) internal shock scenario, focusing on the impact of the collision dynamics between two shells on the fireball evolution.
	In addition to the inelastic case, in which plasma shells merge when they collide, we study the Ultra Efficient Shock scenario, in which a fraction of the internal energy is re-converted into kinetic energy and, consequently, the two shells survive and remain in the system. 
	We find that in all cases a quasi-diffuse neutrino flux from GRBs at the level of $10^{-11}$ to $10^{-10} \, \mathrm{GeV \, cm^{-2} \, s^{-1} \, sr^{-1}}$ (per flavor) is expected for protons and a baryonic loading of ten, which is potentially within the reach of IceCube-Gen2. The highest impact of the collision model for multi-messenger production is observed for the Ultra Efficient Shock scenario, that promises high conversion efficiencies from kinetic to radiated energy. However, the assumption that the plasma shells separate after a collision and survive as separate shells within the fireball is found to be justified too rarely in a multi-collision model that uses hydrodynamical simulations with the \textsc{PLUTO} code for individual shell collisions.
\end{abstract}
\received{}
\accepted{}

\section{Introduction}

Gamma-Ray Bursts (GRBs) have been proposed as plausible candidates for the origin of the Ultra-High Energy Cosmic Rays (UHECRs) and neutrinos, invoking photohadronic interactions in the fireball scenario~\citep{Waxman:1997ti}. It is however evident from GRB stacking searches that GRBs cannot be the dominant source of the observed diffuse astrophysical neutrino flux~\citep{Abbasi:2012zw,Aartsen:2017wea}. In radiation models of high-luminosity GRBs where all emission regions look alike (one-zone models), the nominal predictions for the neutrino flux are in tension with these stacking searches in spite of recently improved neutrino flux estimates~\citep{Hummer:2011ms,Li:2011ah,He:2012tq}. Dedicated scans of the source parameters (including the baryonic loading, the luminosity in protons versus gamma-rays) and fits to the UHECR spectrum and composition data confirm that the simple one zone emission picture is in tension with neutrino data for most of the parameter space~\citep{Baerwald:2014zga, Biehl:2017zlw}. 

As a possible solution multi-collision models have been proposed, which tend to predict a lower neutrino flux~\citep{Globus:2014fka,Bustamante:2014oka,Bustamante:2016wpu} for the same baryonic loading. These models also demonstrate that the different messengers (neutrinos, cosmic rays, gamma-rays) are predominantly emitted at different radii within the GRB jet~\citep{Bustamante:2014oka}. Also the stochasticity pattern and the time delays between different energy bands in the light curves can contain information on the neutrino production efficiency~\citep{Bustamante:2016wpu}.

As specific implementations of the fireball model~\citep{Rees:1992ek,Rees:1994nw}, most GRB multi-collision models invoke emission from internal shocks ~\citep{Kobayashi:1997jk,Daigne:1998xc}, in which a set of plasma shells is emitted from an intermittent central engine
with a specific distribution of Lorentz factors. The emitted shells can have, for instance, equal masses or equal energies, which are related through the relation $E_{\mathrm{kin}} = \Gamma \, m$. Because of the different velocities, the shells will eventually catch up with each other and collide inelastically, forming a merged shell that continues to propagate through the jet. In each shell collision, shocks form that accelerate particles, converting internal energy (from the inelastic collision) into non-thermal radiation.

In principle, multi-collision models provide a natural explanation for the the fast time variability and variety of light curve shapes observed in GRBs; see \eg\ \citet{Kobayashi:1997jk,Daigne:1998xc,Spada:1999fd,Beloborodov:2000nn,Kobayashi:2001iq,Daigne:2003tp}. A fundamental problem is, however, the moderate dissipation efficiency of kinetic energy into non-thermal particles. On one hand, the afterglow emission caused by the shells running into the circumburst medium after the prompt phase might be too high if a large amount of energy remains in the jet. One the other hand extremely powerful engines are required to reach the observed gamma-ray luminosities. The {\em Ultra Efficient Shock scenario} has been proposed as a potential solution to this problem~\citep{Kobayashi:2001iq}, in which a fraction of the internal energy is re-converted into kinetic energy. Consequently, the two shells survive and remain in the system after the collision, resulting in swift thermalization. In this scenario a high overall dissipation efficiency can be achieved, even if the efficiency of each individual two-shell collision is low.
A key ingredient of the scenario is the assumption that two shells remain after each collision, whereas hydrodynamical studies and analytical estimates \citep{Kino:2004uf} demonstrate that this assumption is depending on the collision parameters and that most collisions likely result in only one merged shell.

In this work, we revisit the multi-collision multi-messenger models in \citet{Bustamante:2014oka,Bustamante:2016wpu} from the point of view of the collision dynamics and study the impact on the production of multiple messengers. We employ the methods from \citet{Baerwald:2011ee,Hummer:2011ms,Biehl:2017zlw} for the radiation calculations and use broken power-law target photon spectra that resemble observations. After recapping the previous approach, \ie{},  the fully inelastic case with equal mass injection in \Sec~\ref{sec:Reference_model}, we scan the parameter space for the two-shell collision for configurations with two post-collision shells, and scrutinize the plausability of this assumption in the fireball evolution. The multi-messenger production is studied in \Sec~\ref{sec:alternative_models} for alternative collision models: A) a version of the reference model, in which a fraction of the internal energy is converted into adiabatic expansion of the merged shell; B) the ideal Ultra Efficient scenario according to \citet{Kobayashi:2001iq}; and C) a ``hybrid'' model, where \textsc{PLUTO} is used to determine the fate of each two-shell collision individually. We discuss the impact on the observables (light curves and neutrino fluxes) in \Sec~\ref{sec:multimessenger} and conclude in  \Sec~\ref{sec:summary}.

\section{Methods and reference model}
\label{sec:Reference_model}
As the reference case, we pick the collision model from \citet{Bustamante:2016wpu}, based on \citet{Kobayashi:1997jk} and recapitulate the main ideas and mathematical expressions before entering the discussion of alternatives in the next sections. 

The relativistic outflow of the jet is discretized as a one-dimensional sequence of spherical plasma shells with different velocities and masses. When a faster shell catches up with a slower one, they interact and collisionless shocks convert kinetic energy into internal energy (which will be radiated as non-thermal particles). During the interaction, the two shells merge into a single new shell which will continue to propagate as a part of the jet (see \citet{Bustamante:2016wpu} and first subsection for a detailed description). The internal energy remaining from the collision is converted into radiation through the interaction of accelerated particles, based on the model from an updated version of the \textsc{NeuCosmA} software \citep{Biehl:2017zlw}.\footnote{The update impacts the description of the optically thick (to photo-hadronic interactions) case, see App.~C of \citet{Biehl:2017zlw} for details. This leads to slightly lower predicted cosmic ray fluxes at the highest energies. The baryonic loading is defined with respect to the injection luminosity (previously, the steady state density), resulting in a more transparent evaluation of the energy budget.}

In this section, the ``GRB~1'' in \citet{Bustamante:2016wpu} acts as a reference parameters set, but with an equal-mass instead of equal-energy setup, resulting in a few quantitative differences, which are discussed in \App~\ref{app:equal_energy}. It will become clear in the next sections why this choice allows for a better comparison with the alternative models. For the sake of simplicity, we focus on the proton-only case in this study. 
Changing the composition would not affect the level of the neutrino flux, as it mainly depends on the energy injected per nucleon. However the maximal energy of neutrino flux is lower for heavier injection, as the maximal rigidity is reduced by $Z / A$. See \citet{Biehl:2017zlw} for a detailed discussion of nuclear injection in GRBs.

\subsubsection*{Collision model}

A plasma shell is characterized by its mass ($m$), width ($l$) and the Lorentz factor ($\Gamma$) with respect to the engine frame. When a fast (``rapid'', index $r$) shell collides with a slow (index $s$) shell, forward and reverse shocks develop. They propagate from the contact discontinuity (''CD'', which is the surface separating the initial slow from the fast shell) through the joint density profile initially compressing the matter (see \figu{hydro_examples} for an illustration of the setup). After the shock waves reach the edges of the density profile, two rarefaction waves develop that propagate back towards the CD. They decompress the matter and reconvert internal energy into kinetic energy. The kinetic properties of the $k$-th shell are determined by the Lorentz factor $\Gamma_k$, the mass $m_k$ or the kinetic energy $E_{\text{kin},k}$, the radius $R_k$ and the width $l_k$. The other properties, such as the speed $\beta_k$, the volume $V_k = 4 \, \pi \, R_k^2 \, l_k$ and the density $\rho_k$, can be derived. In the following, all unprimed quantities (unless explicitly given in the observer's frame) are given in the engine frame, and all primed quantities refer to the shock rest (or merged shell) frame.

The simulation of the reference model initially contains 1000 shells with their Lorentz factors $\Gamma_k$'s sampled from a log-normal distribution
\begin{align}
    \ln \left( \frac{\Gamma_{k} - 1}{\Gamma_0 - 1} \right) = A \cdot x \, ,
    \label{equ:gammadis}
\end{align}
where $x$ is distributed as a Gaussian $P(x) dx = \exp(-x^2)/\sqrt{2 \pi} dx$ with $\Gamma_0 = 500$ and $A = 1.0$. 
We further assume that the spatial separation for all shells is equal to their width ($l_k = L_k = 0.01\ \mathrm{c \ s } $, equivalent to the temporal separation $\delta t_k= 0.01\ \mathrm{s}$ for relativistic shells) and that the innermost shell starts at a radius $R_\text{min}$. The initial distribution of kinetic energies $E_{\text{kin}, k}$ is a free model choice, and typical assumptions are equal shell masses $m_k$, energies $E_{\text{kin},k}$ or densities $\rho_k$. As previously mentioned, the reference model uses the equal-mass case. From this initial setup the system evolves until all shells have reached the circumburst medium at $R_\text{max}$.

For the collision between a {\em rapid} and a {\em slow} shell the properties for the {\em merged} shell (index $m$) are determined from simple analytical expressions following \citet{Kobayashi:1997jk}. The Lorentz factor $\Gamma_m$ follows from the conservation of momentum as 
\begin{align}
	\label{equ:merged_gamma}
    \Gamma_m = \sqrt{ \frac{m_r \Gamma_r + m_s \Gamma_s}{m_r / \Gamma_r + m_s / \Gamma_s} } 
\end{align}
(using the approximation $\beta = 1 - 1/(2\Gamma^2)$ that holds for $\Gamma \gg 1$).
The internal energy available for radiation is then the difference of kinetic energies before and after the (inelastic) collision:
\begin{align}
    \label{equ:ediss_multicoll}
    E_{\text{int},m} = m_r \Gamma_r + m_s \Gamma_s - (m_r + m_s) \Gamma_m \,.
\end{align}
We define the efficiency of each collision as the ratio between the dissipated energy and the produced internal energy:
\begin{align}
    \label{equ:eta_def}
    \eta = \frac{E_{\text{diss},m} }{E_{\text{int},m}}
\end{align}
The idealized assumption of $\eta = 1$ in the {\em Reference} model will be modified for the alternative models in \sect{alternative_models}.

The Lorentz factors of the forward (fs) and the reverse shock (rs) are determined from
\begin{align}
    \label{equ:shock_velocities}
    \Gamma_{\mathrm{fs \, (rs)}} = \Gamma_m \cdot \sqrt{ \frac{1 + 2 \Gamma_m / \Gamma_{s \, (r)}}{2 + \Gamma_m / \Gamma_{s \, (r)}} } \,.
\end{align}

The timescale of emission $t_{\mathrm{em}}$ is estimated from the time taken by the reverse shock to cross the rapid shell
\begin{align}
	\delta t_{\mathrm{em}} = \frac{l_r}{\beta_{r} - \beta_{\mathrm{rs}}} \,.
\label{equ:emission_timescale}
\end{align}
This timescale is observed Doppler-boosted in the observer's frame. The width of the compressed merged shell is computed from 
\begin{align}
    l_m = l_s \frac{\beta_{\mathrm{fs}} - \beta_m}{\beta_{\mathrm{fs}} - \beta_s}
     + l_r \frac{\beta_{m} - \beta_{\mathrm{rs}}}{\beta_{r} - \beta_{\mathrm{rs}}} \, . 
\label{equ:shell_compression}
\end{align}

\revise{For \equ{shock_velocities} to \equ{shell_compression} we follow the description given in \citet{Kobayashi:1997jk}. This formulation might not be valid in the non-relativistic regime, however collisions in the non-relativistic regime dissipate little energy 
and therefore contribute only marginally to the multi-messenger emission.
Even though $t_{\mathrm{em}}$ and $l_m$ might be slightly miss-estimated for those collisions, the impact on our main results is therefore small. }

\subsubsection*{Radiation model}

The collision parameters derived in the previous section are now used for the computation of the energy density and the emission spectra in the merged shell (primed frame). It is assumed that fractions of the injected luminosity $L'_{\text{diss},m} \simeq E'_\text{diss} \, c/l'_m$ are distributed into protons ($\epsilon_p$), electrons ($\epsilon_e$), and the magnetic field ($\epsilon_B$), such that $\epsilon_p+\epsilon_e+\epsilon_B=1$. Normally it is assumed that the non-thermal electrons loose energy quickly via synchrotron radiation, implying that gamma-rays will carry a comparable amount of energy $\epsilon_\gamma \simeq \epsilon_e$. We assume a baryonic loading of $\epsilon_p/\epsilon_\gamma = 10$. It is convenient to relate these quantities to the (observed gamma rays), consequently $\epsilon_\gamma = \epsilon_B = 1/12$ and $\epsilon_p = 10/12$.
 
We assume a power-law spectrum motivated by Fermi shock acceleration ($\propto (E'_p)^{-2} \exp(-E'_p / E'_{p,\text{max}})$) for the injected proton component. The maximal energy $E'_{p,\text{max}}$ is found by comparing the timescales of efficient acceleration $t'^{-1} \simeq  c/R'_L$ ($R'_L$ is the Larmor radius), photohadronic and adiabatic energy losses (for a detailed discussion of the maximum energies see also \citet{Samuelsson:2018fan}). The proton spectrum is normalized to the available luminosity fraction ($\epsilon_p$) as outlined above.

For the target photon spectrum, we do not perform explicit self-consistent radiation calculations, but instead assume a shape motivated by observations-- a broken power law with spectral indices $-1$ and $-2$ below and above the break energy $\varepsilon'_\text{break} \simeq 1\ \mathrm{keV}$, respectively \cite{Gruber:2014iza}. The maximal photon energy is assumed to be limited to $\varepsilon'_\text{max} = 10^{6}\ \mathrm{GeV}$ and reduced in case $\gamma \gamma$-annihilation sets in at lower energies. The normalization is performed equivalent to the proton case with the fraction $\epsilon_\gamma$. With these power-law indices, the energy densities depend (at most) logarithmically on the maximal and minimal energies, reducing the impact of our choice of $\varepsilon'_\text{max}$.
Successful  modeling of GRB prompt spectra within the internal shock model has been performed in e.g.\ \cite{Bosnjak:2008bd, Daigne:2010fb, Bosnjak:2014hya}. Realistic values for the synchrotron peak energy and spectral indices can be achieved, for instance, by allowing a small fraction of electrons to be accelerated to high energies (as in \cite{Eichler:2005ug, Spitkovsky:2008fi}). 
However, an explicit modeling of photon synchrotron spectra is not within the scope of this work.

The coupled proton-neutron system is evolved to the steady state using \textsc{NeuCosmA} \citep{Biehl:2017zlw}, which takes photohadronic, pairproduction, adiabatic and synchrotron losses into account. The neutrons escape freely, whereas the protons are assumed to be magnetically confined and to escape only from the boundaries (within their Larmor radius), referred to as ``direct escape''~\citep{Baerwald:2013pu}. This assumption implies that close to the photosphere the cosmic ray emission is dominated by neutron escape, and for large collision radii by direct proton  escape \citep{Bustamante:2014oka}. The mechanism for cosmic ray escape is currently discussed, and it is likely that several competing components contribute; see discussion in~\citet{Zhang:2017moz}. The implementation for secondary particle emission is described in great detail in \citet{Biehl:2017zlw}. 

The model does not account for the emission from sub-photospheric collisions, for which the optical thickness to Thomson scattering is larger than one --  resulting in a different, thermalized shape of the target photon spectra. Even if cosmic-ray acceleration could take place at such low radii, the high radiation densities will prevent the particles from reaching the UHE range. Since the pion production efficiency scales with the density similar to the Thomson optical depth, the neutrino production is most efficient close to the photosphere. There could be a significant contribution from sub-photospheric collisions; hence, our neutrino flux estimate shall be considered as minimal prediction. We chose examples in which the fraction of sub-photospheric collisions is small to reduce the impact of this effect. 

\subsubsection*{Discussion of Reference model}

\begin{figure*}[tb]
	\centering
	\makebox[\textwidth][c]{
		\subfloat[(a)]{\includegraphics[width=.35 \textwidth]{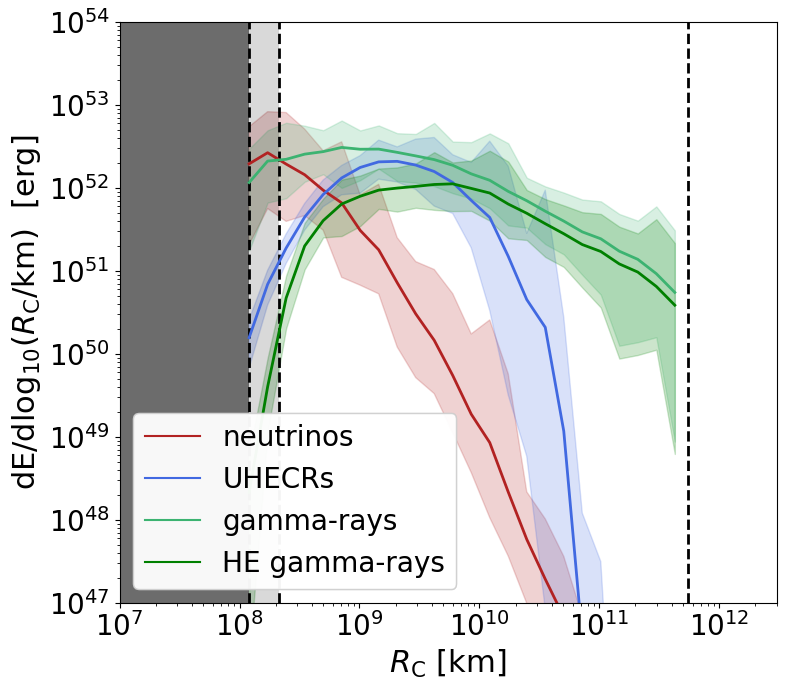}}
		\subfloat[(b)]{\includegraphics[width=.35 \textwidth]{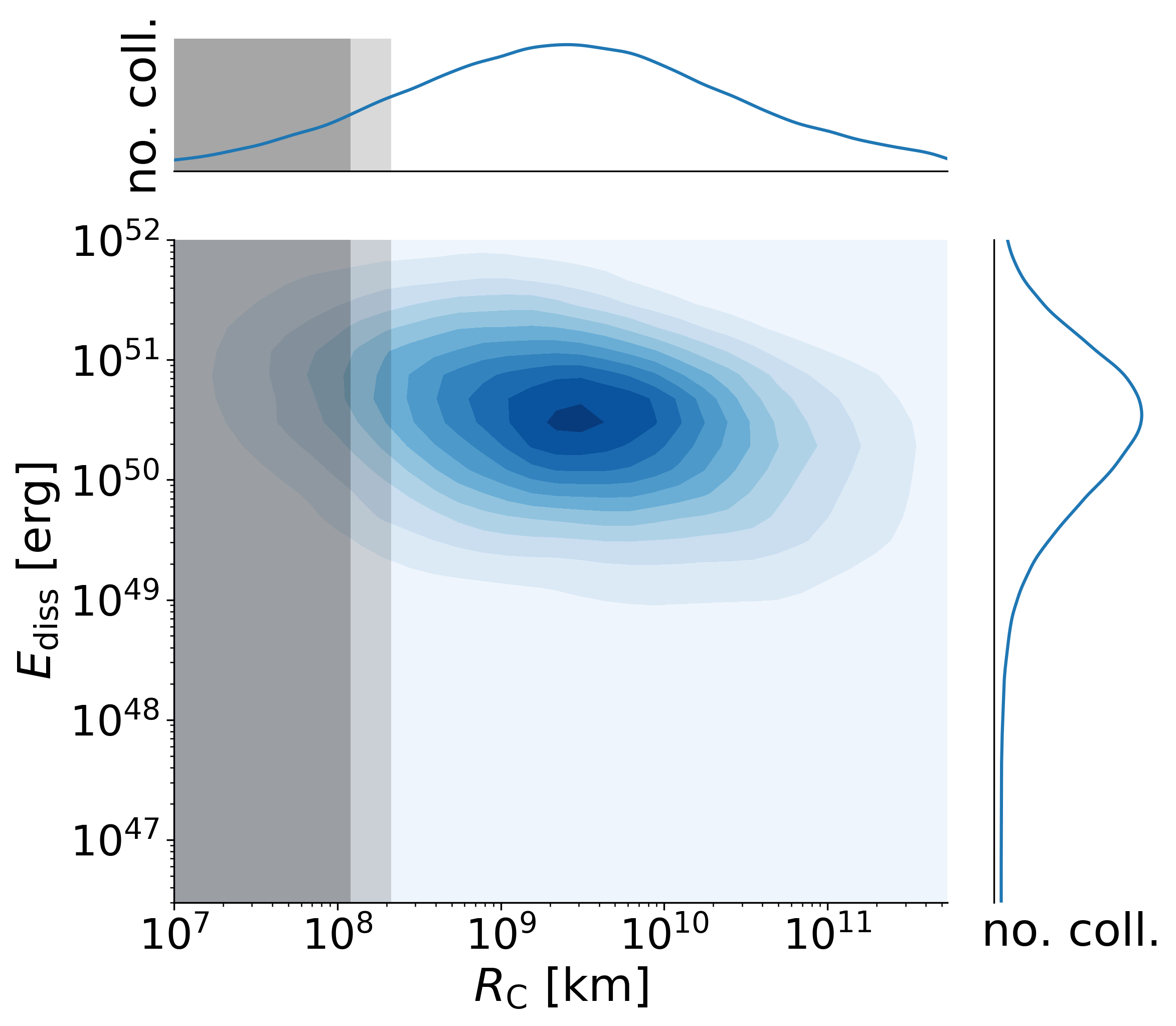}}
		\subfloat[(c)]{\includegraphics[width=.35 \textwidth]{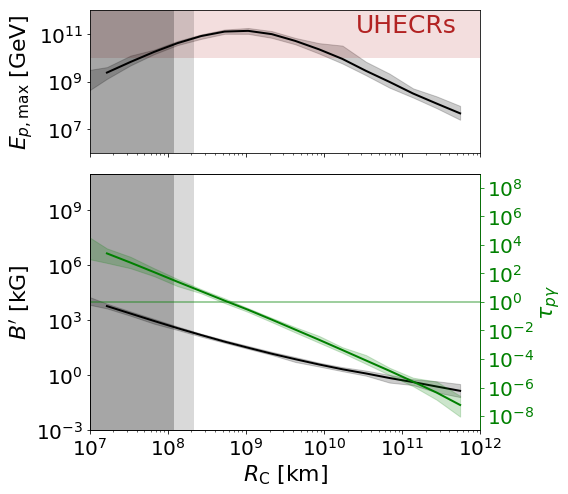}}
		  
 	}
	\caption{\label{fig:panel_ref} Reference model, corresponding to GRB~1 from \citet{Bustamante:2016wpu} for $E_{\gamma, \mathrm{iso}} = 5.2  \cdot 10^{52}\ \mathrm{erg}$ but for a constant mass injection rate: 
	(a) Energy dissipated in neutrinos, high-energy (HE) gamma rays, gamma-rays and UHECR as a function of collision radius. We define ``HE gamma rays'' as gamma rays that have energies above 1~GeV in the source frame, and UHECRs as cosmic rays that have energies above $10^{10} \, \mathrm{GeV}$. 
	In the dark gray-shaded area only sub-photospheric collisions occur, while in the light gray-shaded both sub- and super-photospheric collisions occur; subphotospheric collisions are not included in this model.
	(b) The number of collisions as a function of collision radius and dissipated energy  (the contour plot shows the density of the number of collisions). 
	(c) The averaged maximal proton energy (upper panel), as well as the obtained magnetic field and the optical depth to $p\gamma$ reactions (different colors/axes)  as a function of radius (lower panel). All plots have been computed averaging over 100 random seeds for the initial shell distribution. The curves show the average, and the shaded areas the fluctuations.}
\end{figure*}

We will use the format of \figu{panel_ref} to characterize the behavior of different models, and hence explain it here in greater detail. Since the initial configuration of the 1000 shells is drawn from the distribution in \equ{gammadis}, the model naturally produces stochastic fluctuations in the variables. We compute 100 representations if not otherwise noted. In the figures, we show the average as a solid curve and the region between the edge cases with a shaded band. 

The left panel shows the differential energy dissipation in the engine frame for the different messengers: neutrinos, UHECRS above $10^{10} \, \mathrm{GeV}$, gamma-rays, and gamma-rays above $1 \, \mathrm{GeV}$ only. In the dark-shaded region, all collisions occur below the photosphere, in the light-shaded region, the collision may be above or below the photosphere (depending on the other shell parameters). The result is consistent with earlier works \citep{Bustamante:2014oka, Bustamante:2016wpu}: neutrinos originate near the photosphere due to the high density; UHECRs prefer intermediate collision radii since high magnetic fields are  required for efficient acceleration -- but not as high as that radiation losses (such as synchrotron radiation) limit the maximal energy; gamma-rays trace the region where most energy is dissipated (see middle panel and discussion below); HE gamma-rays above 1~GeV in the source frame prefer larger collision radii where maximal energies are not dominated by losses. The different astrophysical messengers originate from different radii of the same jet, thus drawing attention to the collision model that defines the connections among these different regions. The statistical fluctuations from the initial shell setup (shaded areas) imply some variability in this picture, but they do not change it qualitatively.

The middle panel of \figu{panel_ref} shows the distribution of radius and dissipated energy of the individual collisions in arbitrary units. This figure demonstrates that most of the relevant collisions happen above the photosphere and there is a not very noticable anti-correlation between $R_C$ and $E_{\mathrm{diss}}$, confirming the relation to the gamma-ray output in the left panel.

The maximal cosmic ray energy in the upper right panel of \figu{panel_ref} exhibits a similar shape as the UHECR curve in the left panel, that only accounts for cosmic rays with $E_{\text{p,max}}$ within the red-shaded area.
The lower panel shows the magnetic field $B'$ (left axis, black curve) and the optical thickness to photohadronic interactions evaluated at $E_{\text{p,max}}(R_C)$ (right axis, green curve). Both scale with the collision radius, $\tau_{p \gamma} \sim R_C^{-2}$ and $B' \sim R_C^{-1}$. The collisions close to the photosphere are optically thick to photohadronic interactions, and are responsible for most of the neutrino production (see left panel). 

We define the overall energy dissipation efficiency from kinetic energy to radiation as 
\begin{align}
 \label{equ:epsilon_def}
 \epsilon \equiv E^{\mathrm{tot}}_{\mathrm{diss}}/E^{\mathrm{tot}}_{\mathrm{kin}}
\end{align}
In contrast to $\eta$ it describes the efficiency of the whole system and is an output instead of an input parameter. For the {\em Reference} model we find $\epsilon \approx 36\% $, \ie{}, somewhat higher than for the equal-energy setup in \citet{Bustamante:2014oka,Bustamante:2016wpu}. The reason for this is that the efficiency for individual collision $\eta$ is higher in the equal mass case \citep{Kino:2004uf}.
Compared to the equal-energy setup, the {\em Reference} model discussed here relatively efficiently converts the kinetic energy drawn from the GRB engine into secondaries. We discuss this issue  in greater detail in \sect{alternative_models} for alternative model assumptions, and we compare to the literature in \sect{eff}. The kinetic energies assumed in our models (see \Tab~\ref{tab:model_results}) are at the higher end of observations \citep{Gruber:2014iza}, motivated by the higher baryonic loading required to reach the UHECR energy density observed at Earth.

\section{Probability for two-shell final states}
\label{sec:hydrodynamics}

\begin{figure*}
	\centering
		\subfloat[(a)]{\includegraphics[width=0.9\columnwidth]{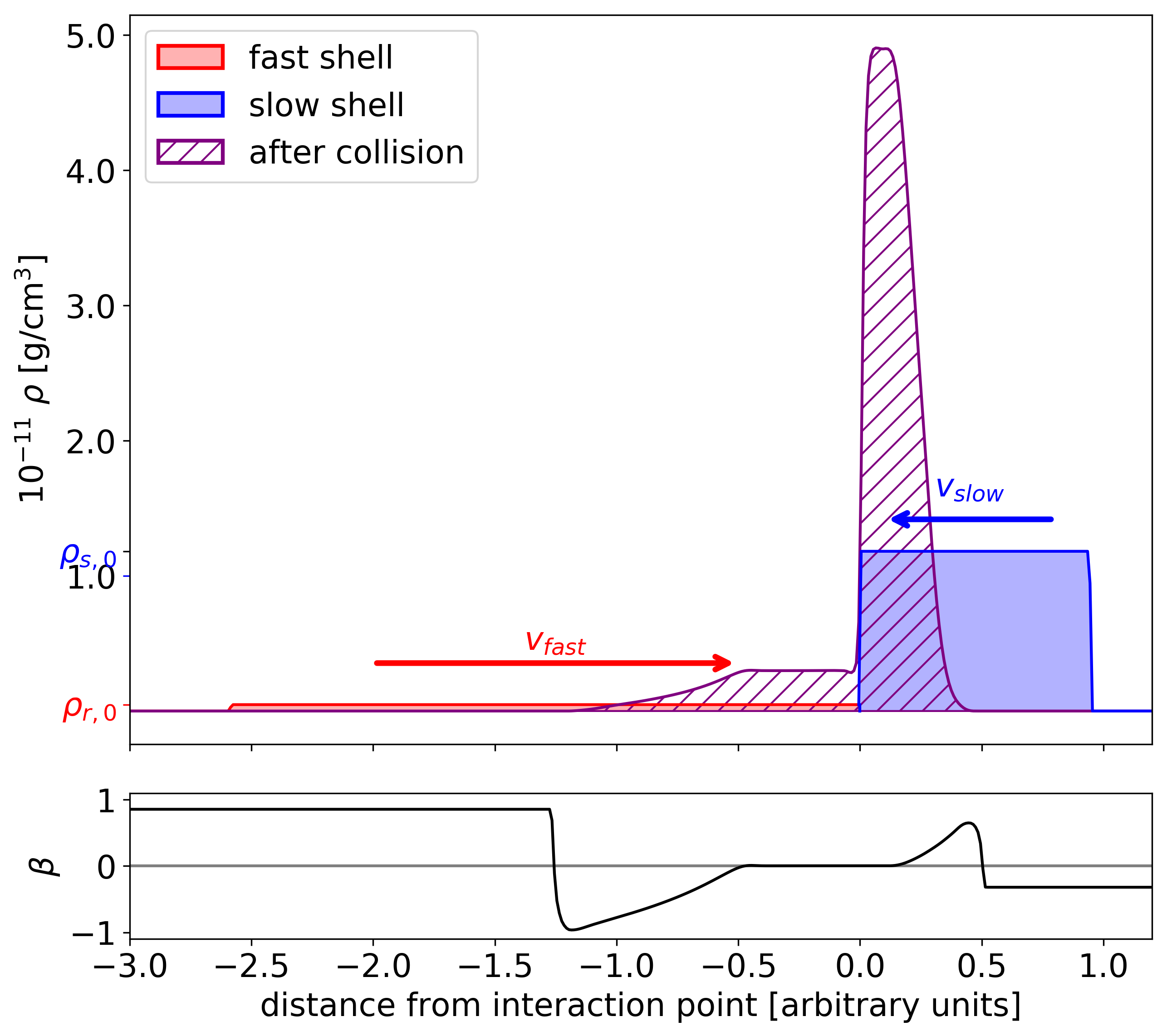}} \hspace*{0.1\columnwidth}
		\subfloat[(b)]{\includegraphics[width=0.9\columnwidth]{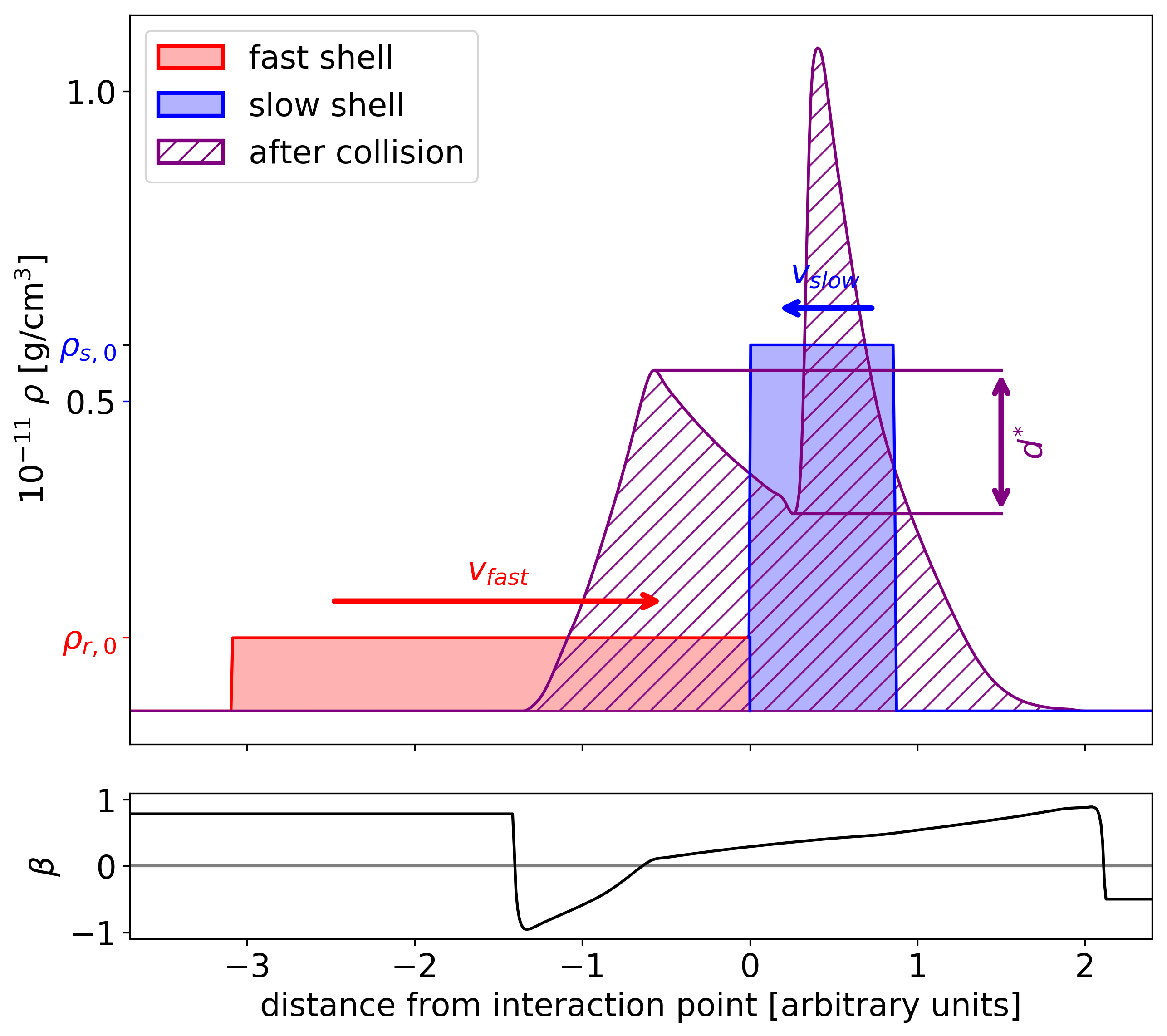}}
	\caption{\label{fig:hydro_examples}
 Collision of a rapid ($r$) and slow ($s$) shell in frame of the contact discontunuity (CD) in the case of equal initial energies (a) or equal masses (b). The mass density profiles at $t^\prime=0$ are in red and blue and the hatched common profile is a snapshot at $t^\prime_{\mathrm{shock}}$ after the collision. The velocity profile $\beta$ of the hatched surface is shown in the lower panels in units of $c$. The Lorentz factors for the  initial shells are $\Gamma_{\mathrm{r}} = 500$ and $\Gamma_{\mathrm{s}} = 100$ and the widths are equal in the source frame ($l_s = l_r$). The snapshot (after the collision) time $t^\prime_{\mathrm{shock}}$ is defined when the reverse and forward shocks have crossed the respective shell. The parameter $d^* =  \rho_{\mathrm{Max}} - \rho_{\mathrm{Min}}$ is the depth of the density dip (where $\rho_{\mathrm{Max}}$ the density at the lower peak) which will be used later as a discriminator between single- and double-peaked profiles.}
\end{figure*}

The key ingredient of the the {\em Ultra Efficient} model by \citet{Kobayashi:2001iq} is the emergence of two shells after a collision. Here, we study the probability of such events using ranges of collision parameters from stochastic GRB multi-collision models with  hydrodynamical simulations, see \appr{PLUTO_methods} for details. 
We use the \textsc{PLUTO} code \citep{Mignone:2007iw} with a one-dimensional setup.
A study for a two-dimensional setup demonstrated qualitatively similar behavior to one dimension \citep{Mimica:2004cv}. Magnetic fields are neglected. The colliding shells are assumed to be cold. For a treatment of arbitrarily hot plasma shells see \eg\ \citet{Peer:2016irw}, who find a possible suppression of the shock formation and a dependence of the energy per particle after the collision on the pre-collision plasma temperature. 
 
The collision process and the post-collision shell configurations are illustrated in \figu{hydro_examples}. Panel~(a) demonstrates the equal energy case (in the source frame) and panel~(b) the equal masses case. In both panels the shells are shown at $t^\prime =0 $, the slow shell is depicted in blue, the fast one in red. The mass density profiles at $t^\prime_{\mathrm{shock}}$, when both shocks have crossed the respective shells, are shown in purple. 
In \figu{hydro_examples}~(a), the resulting mass density profile is clearly single-peaked, while in \figu{hydro_examples}~(b) two distinct peaks moving at different speeds can be seen. 
In order to identify two-shell post-collision configurations, we evaluate the mass density profile at the time when both shocks have crossed the shells and the internal energy is maximal. For a given snapshot in time, the mass density can exhibit multiple peaks; \citet{Kino:2004uf} predict up to three peaks. Theoretical estimates comparing the time scales of the wave propagation (when the shock and rarefaction waves cross the two shells) may result in unrealistic approximations, since double-peaked profiles can rapidly evolve into single-peaked ones. After $t^\prime_{\mathrm{shock}}$, the density profile continues to evolve since the velocity profile is not uniform across the shell(s) (\cf{}, lower panels of \figu{hydro_examples}). Instead, the shell edges move in opposite directions in the CD frame. This leads to a dilution of the density profile as time evolves, invalidating the assumption of a constant shell width for the entire duration of a simulation (see also \citet{Peer:2016irw}). As discussed below, the dissipation of internal energy reduces this effect by slowing down the thermal expansion that ultimately leads to a washed-out single-peaked profile.

For the classification of a two-shell collision, we define the relative depth of the dip between two mass peaks
\begin{equation}
	d = \frac{d^*}{\rho_{\mathrm{Max}}}= \frac{\rho_{\mathrm{Max}} - \rho_{\mathrm{Min}}}{\rho_{\mathrm{Max}}}\ , 
	\label{equ:d_pluto}
\end{equation}
where $\rho_{\mathrm{Min}}$ is the density at the dip between the two maxima and $\rho_{\mathrm{Max}}$ the density at the lower peak (illustration see \figu{hydro_examples}~(b)). 

\begin{figure*}
	\centering
	\makebox[\textwidth][c]{
		\subfloat[(a)]{\includegraphics[width=.35 \linewidth]{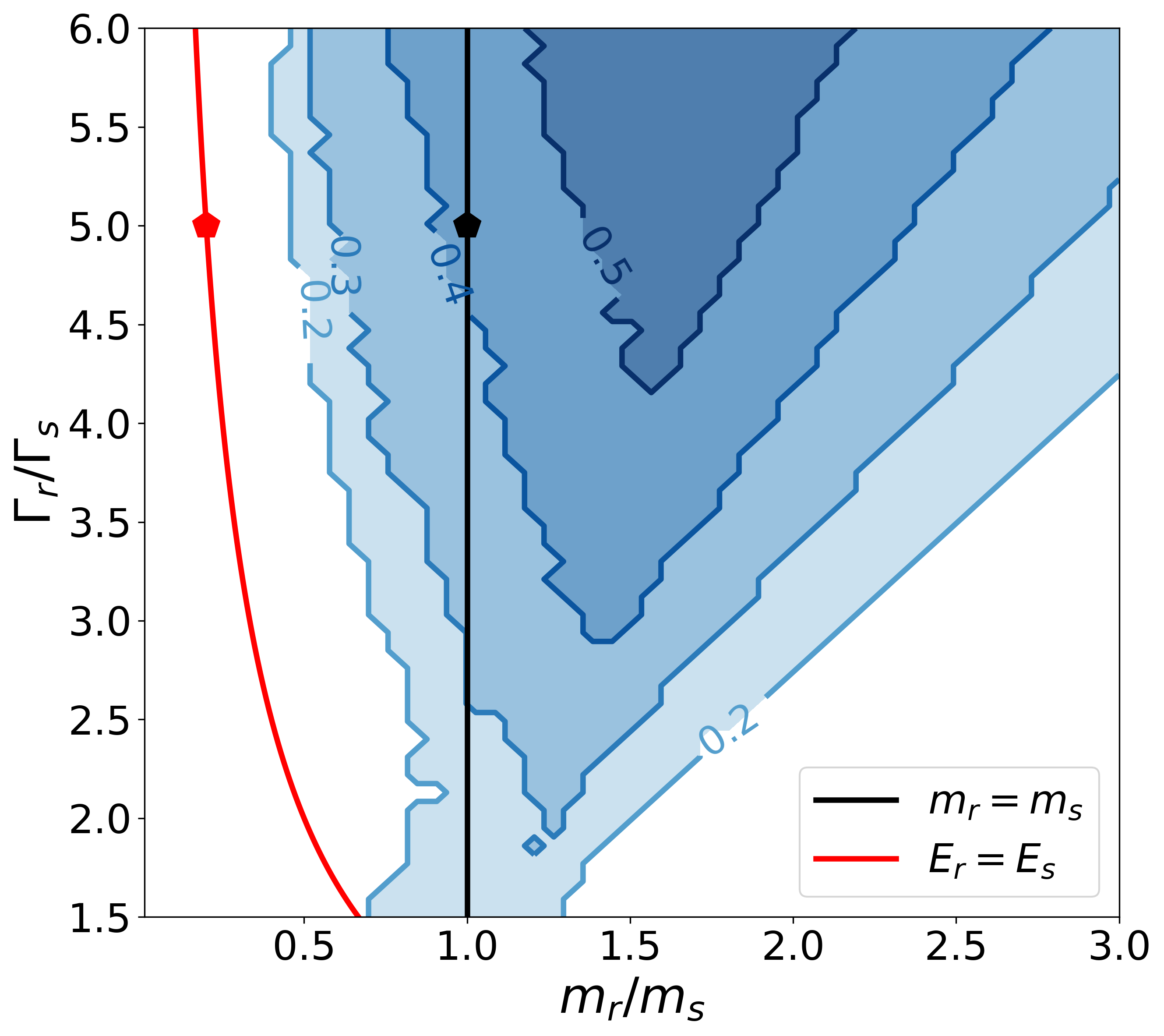}}
		\subfloat[(b)]{\includegraphics[width=.35 \linewidth]{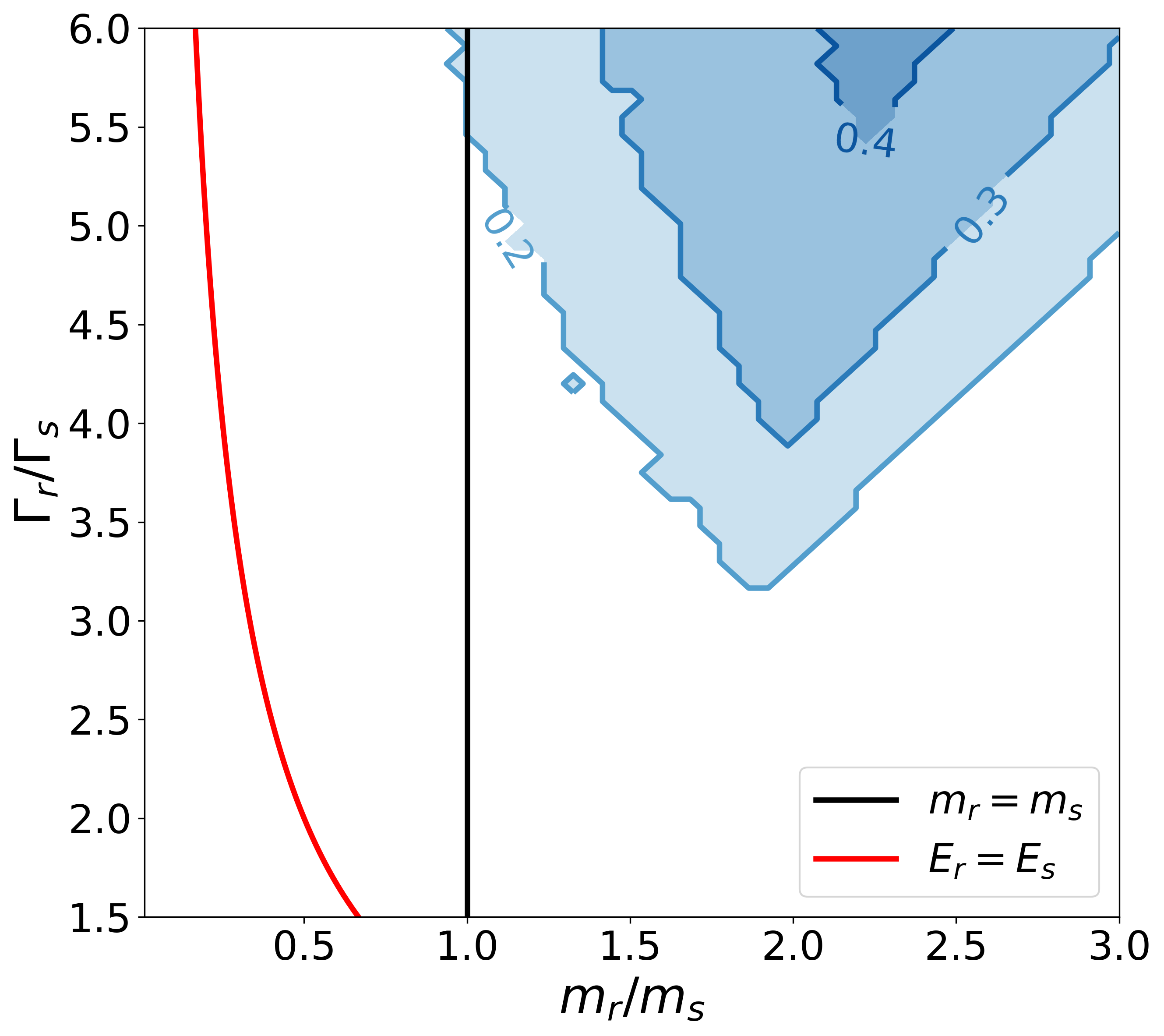}}
		\subfloat[(c)]{\includegraphics[width=.35 \linewidth]{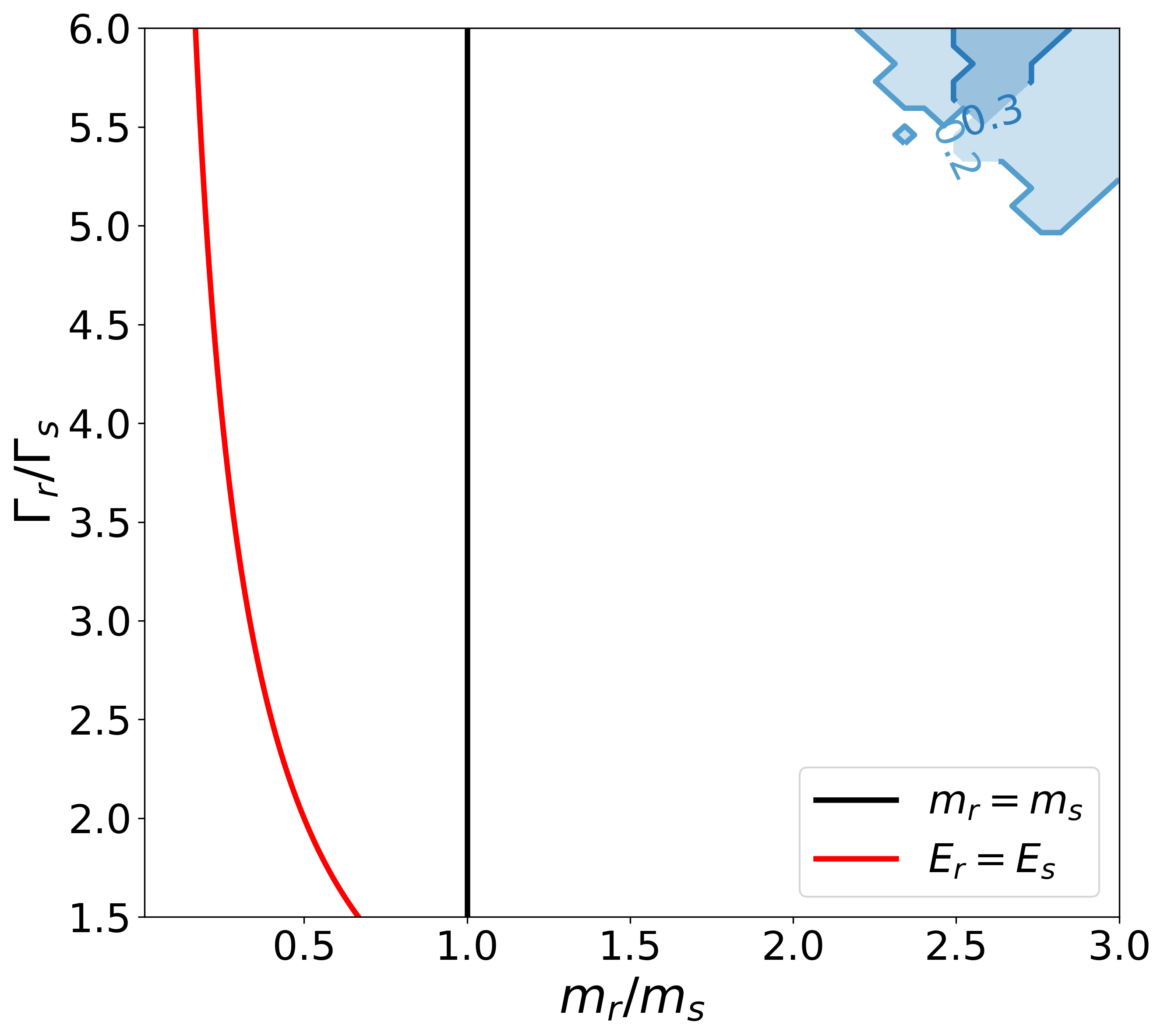}}
	}
	\caption{Relative depth of the dip $d$ between the two shells after the collisions (see \equ{d_pluto}) as a function of the pre-collision mass ratio ($m_\mathrm{r} / m_\mathrm{s}$) and the ratio of Lorentz factors ($\Gamma_{\mathrm{r}} / \Gamma_{\mathrm{s}}$), in all cases we fixed $\Gamma_{\mathrm{s}} = 100$ and $l_{\mathrm{r}}= l_{\mathrm{s}}$. The system is evolved until both shocks have crossed the respective shells. For panel (a) energy dissipation in not included, in panel (b) $20 \%$ of the (theoretically) produced internal energy is removed from the system, in (c) $50 \%$. The red (black) curve corresponds to the case where the colliding shells share the same initial kinetic energy (rest mass). The dots represent the cases shown in \figu{hydro_examples}. 	\label{fig:parameterscan}
	}

\end{figure*}

The main impact on the post-collision mass density profile comes from the pre-collision mass ratio ($m_\mathrm{r} / m_\mathrm{s}$), the Lorentz factors ($\Gamma_{\mathrm{r}} / \Gamma_{\mathrm{s}}$) and the shell widths ($l_{\mathrm{r}} / l_{\mathrm{s}}$). We fix the width ratio $l_{\mathrm{r}} /  l_{\mathrm{s}} = 1 $ since in GRB internal shock models (see \eg\ \citet{Kobayashi:2001iq, Kobayashi:1997jk, Daigne:1998xc} and \citet{Globus:2014fka, Bustamante:2016wpu, Bosnjak:2008bd}), plasma shells are ejected at constant time intervals resulting in equal widths. An alternative choice $l_{\mathrm{r}} = 0.1 \ l_{\mathrm{s}}$ is discussed in the Appendix. 

\figu{parameterscan} shows the depth d as a function of $m_\mathrm{r} / m_\mathrm{s}$ and $\Gamma_{\mathrm{r}} / \Gamma_{\mathrm{s}}$ for the parameter ranges  that enclose most of the collisions in our models (see \sect{alternative_models}). As already noticed by \citet{Kino:2004uf}, pronounced dips, and hence double-peaked density profiles, occur for almost equal shell rest masses and for high Lorentz factor ratios $\Gamma_{\mathrm{r}} / \Gamma_{\mathrm{s}}$. The latter can be understood from the shock-/ rarefaction-wave timescales: If $\Gamma_{\mathrm{r}} / \Gamma_{\mathrm{s}}$ is high, then $l_{s}^\prime \ll l_{r}^\prime $. Therefore, the reverse shock takes substantially longer to cross the fast shell than the forward shock to cross the slow one. As a result, the rarefaction wave from the direction of the slow shell has enough time to create a pronounced dip separating the two shells. 
\newline

The radiation model assumes that a fraction of the internal energy is converted into non-thermal electrons and/or ions which leave the system, which effectively cools the system. To study the impact on the collision dynamics we introduce a simplified energy dissipation term in our \textsc{PLUTO} simulations that removes internal energy from the system until a certain threshold has been reached (for more details see \appr{PLUTO_methods}). The rate of this process is assumed to be proportional to the available internal energy. The impact on the occurrence of two-shell configurations is demonstrated by the contours in \figu{parameterscan}~(b) and~(c) for two efficiency choices. 
The dissipation of internal energy reduces the velocity and amplitude of the rarefaction waves, since those are powered from the available internal energy. Too slow rarefaction waves result in shallower dips in the mass density profile (see also Figure \figu{pluto_appendix_singlecoll}), decreasing $d$ and moving up the contours in \figu{parameterscan}. 

For the Ultra Efficient shock scenario, this result means that the dissipation of internal energy into non-thermal particles significantly reduces the probability for two-shell final states. At the same time, the reduction of internal energy available to drive the rarefaction waves increases the lifetime of a two-shell configuration since the thermal expansion of a the double-peaked profile slows down. Hence, a higher energy dissipation rate reduces the emergence probability of two-shell configurations, but increases their lifetime. The model ``PLUTO'' described in the next section takes the first effect more rigorously into account.

\section{Alternative collision models}
\label{sec:alternative_models}

In this section, we define three alternative models that use different assumptions for the collision of two shells, and study the impact on the multi-messenger production. The format of the discussion follows \sect{Reference_model} where the details on the \emph{Reference} model can be found, which is used as benchmark case. 

The \emph{Reduced Efficiency} model assumes that in each collision a fraction $\eta=0.5$ of the internal energy in \equ{ediss_multicoll} is dissipated as radiation, whereas the remaining kinetic energy goes into the adiabatic expansion of a single merged shell; the \emph{Ultra Efficient} model assumes partially inelastic collisions in which the remaining internal energy is re-converted into kinetic energy, always resulting in two shells after the collision~\citep{Kobayashi:2001iq}; in the \emph{PLUTO} model each two-shell collision is simulated individually with the \textsc{PLUTO} code. 

Since each model would produce a different result for the same initial shell setup, we modify the setup of each model to re-produce comparative burst durations, variability times derived from the light curve, and total gamma-ray luminosities. Here, the variability timescale refers to the fastest time variability observed on top of longer-lasting light curve pulses. By construction, all GRBs are normalized to release the same amount of energy in photons in the optically thin regime. The burst duration ist determined by the initial size of the system (the sum of all initial shell widths and separations), which matches among the different models. The time variability primarily depends on the number of collisions for a constant burst duration, which scales with the number of initial shells. However, in the {\em Ultra Efficient} model shells do not merge when colliding, leading to  substantially more collisions for the same number of initial shells. We compensate for this by reducing $N^{\mathrm{shells}}_{\mathrm{initial}} = 1000$ to 125. 
While the resulting variability timescales are rather small, they do not necessarily translate into observed time variabilities, since these are limited by the instrument's response (\ie, the actual variability could be smaller).
We have verified that our results related to the multi-messenger production are not qualitatively affected by the choice of $1000$ initial of shells, see App.~\ref{sec:reduce_number_of_shells}.
In all cases we assume an engine with a constant mass outflow and constant up- and downtimes resulting in equal initial shell widths and separations. As previously discussed, a constant mass outflow is more likely to result in a splitting into two shells, which the \textit{Ultra Efficient} model is based on. An overview of the model parameters is given in \Tab~\ref{tab:modelcomp}.

\begin{table*}[tbp]
	\centering
	\begin{tabular}{ l | r r | c r c}
		\toprule
		\hline
		& \multicolumn{2}{c |}{Initial setup}	& \multicolumn{3}{c}{Single collision result} \\
		Model name & $N^{\mathrm{shells}}_{\mathrm{inital}}$ & $\frac{l^{\mathrm{shells}}_{\mathrm{initial}}}{\mathrm{c}}$& $N^{\mathrm{shells}}_{\mathrm{post-coll.}} $ & $\eta$ & $l^{\mathrm{shells}}_{\mathrm{post-coll.}}$ \\
		\hline
		Reference  		& 1000                  & 0.01s	& 1     & 1      & $ < l_r + l_s$      \\
		Reduced Efficiency    	& 1000                  & 0.01s	& 1     & 0.5    & $ = l_r + l_s$      \\
		Ultra Efficient  	& 125                   & 0.08s	& 2     & 0.5    & $l_r$, $l_s$\\
		PLUTO  			& 1000                  & 0.01s & 1-2	& 0.5    & $ < l_r + l_s$      \\
		\hline
	\end{tabular}
	\caption{Qualitative comparison of the different collision models.
	Here $N^{\mathrm{shells}}_{\mathrm{inital}}$  is the number of shells ejected by the source, $l^{\mathrm{shells}}_{\mathrm{initial}}$ are the initial shell widths (assumed to be equal for all shells) and $N^{\mathrm{shells}}_{\mathrm{post-coll.}}$ is the number of shells produced by a single collision. In each collision, the dissipated energy is given by $\eta \equiv E_{\mathrm{diss}}/E_{\mathrm{int}}$, and the width of the shell(s) after the collision is given by $l^{\mathrm{shells}}_{\mathrm{post-coll.}}$. In the case of the {\em PLUTO} model, the number of post-collision shells as well as their widths are obtained from an analysis of the post-collision mass density profile from the simulation.}
	\label{tab:modelcomp}
\end{table*}
\begin{table*}
	\begin{tabular*}{\textwidth}{l @{\extracolsep{\fill}} | c c c c | c c c}
		\toprule
		\hline
					&$\epsilon$ [\%]  	& $t_{\nu}$ [ms] 	& $N_{\mathrm{coll}} $ 	& $E^{\mathrm{tot}}_{\mathrm{kin}}$ [$10^{54}$ erg]& $E_{p, \mathrm{tot}}^{\mathrm{iso}} / E_{\gamma, \mathrm{tot}}^{\mathrm{iso}}$  & $E_{\nu, \mathrm{tot}}^{\mathrm{iso}} / E_{\gamma, \mathrm{tot}}^{\mathrm{iso}}$ & $E_{p,\mathrm{max}}$ [$10^{12}$ GeV] \\ \hline
		Reference  		&  35.8 $\pm$ 1.4 	&  55.2 $\pm$ 1.3 	&  970.1 $\pm$ 3.3 	&  1.75 $\pm$ 0.07 	&  0.42 $\pm$ 0.03	&  0.29 $\pm$ 0.05	&  1.2 $\pm$ 0.4  \\ 
		Reduced Efficiency	&  17.9 $\pm$ 0.7 	&  54.8 $\pm$ 1.3 	&  976.0 $\pm$ 3.3 	&  3.50 $\pm$ 0.13 	&  0.56 $\pm$ 0.04	&  0.24 $\pm$ 0.05	&  1.2 $\pm$ 0.4  \\ 
		Ultra Efficient 	&  36.0 $\pm$ 4.3	&  47.5 $\pm$ 10.7 	&  1107 $\pm$ 220 	&  1.76 $\pm$ 0.22 	&  0.62 $\pm$ 0.06	&  0.14 $\pm$ 0.06	&  1.2 $\pm$ 0.5  \\ 
		PLUTO	 		&  21.2 $\pm$ 1.4 	&  50.2 $\pm$ 1.5 	&  1055 $\pm$ 9 	&  2.95 $\pm$ 0.20 	&  0.96 $\pm$ 0.4	&  0.18 $\pm$ 0.04	&  1.4 $\pm$ 0.6  \\
		\hline
	\end{tabular*}
	
	\caption{\label{tab:model_results}
		Results for the different models: $\epsilon \equiv E^{\mathrm{tot}}_{\mathrm{diss}}/E^{\mathrm{tot}}_{\mathrm{kin}}$ describes the overall dissipation efficiency of the burst, $t_{\nu}$ the (observable) time variability of the light curve assuming a redshift of $z=2$, $N_{\mathrm{coll}}$ the total number of collisions and $E^{\mathrm{tot}}_{\mathrm{kin}}$ the total initial kinetic energy. The energies $E_{\nu, \mathrm{tot}}^{\mathrm{iso}}$, $ E_{\gamma, \mathrm{tot}}^{\mathrm{iso}}$ and $E_{p, \mathrm{tot}}^{\mathrm{iso}}$ are the total energies released in neutrinos, gamma-rays and UHECRs above $10^{10} \, \mathrm{GeV}$ in the engine frame (where only super-photospheric collisions are taken into account), and $E_{p,\mathrm{max}}$ is the maximal cosmic ray energy achieved in the engine frame. We show the mean value over  all statistical realizations as well as the standard deviation obtained from the Monte Carlo simulation.}
\end{table*}

\begin{figure*}[tbp]
	\centering
	\makebox[\textwidth][c]{
		\subfloat{\includegraphics[width=.35 \textwidth]{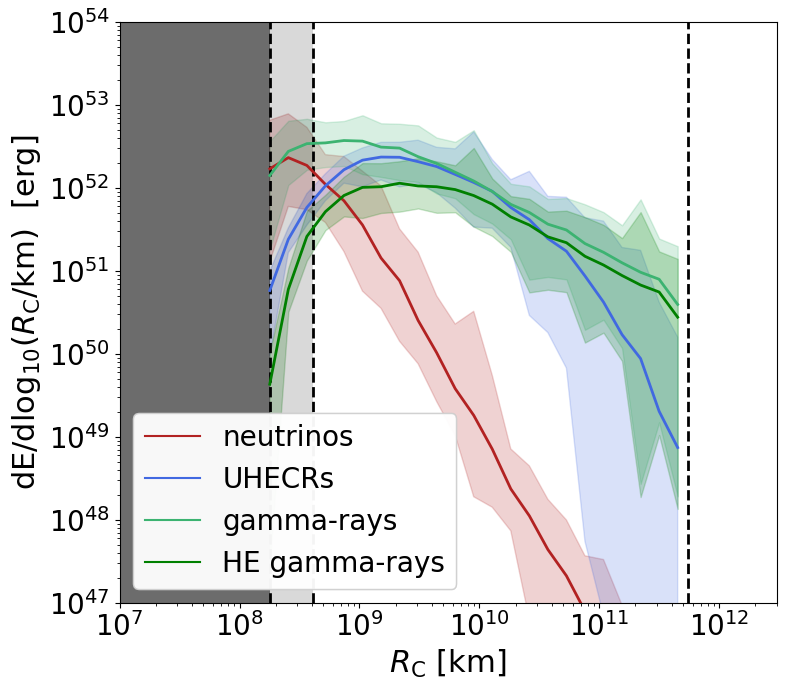}}
		\subfloat{\includegraphics[width=.35 \textwidth]{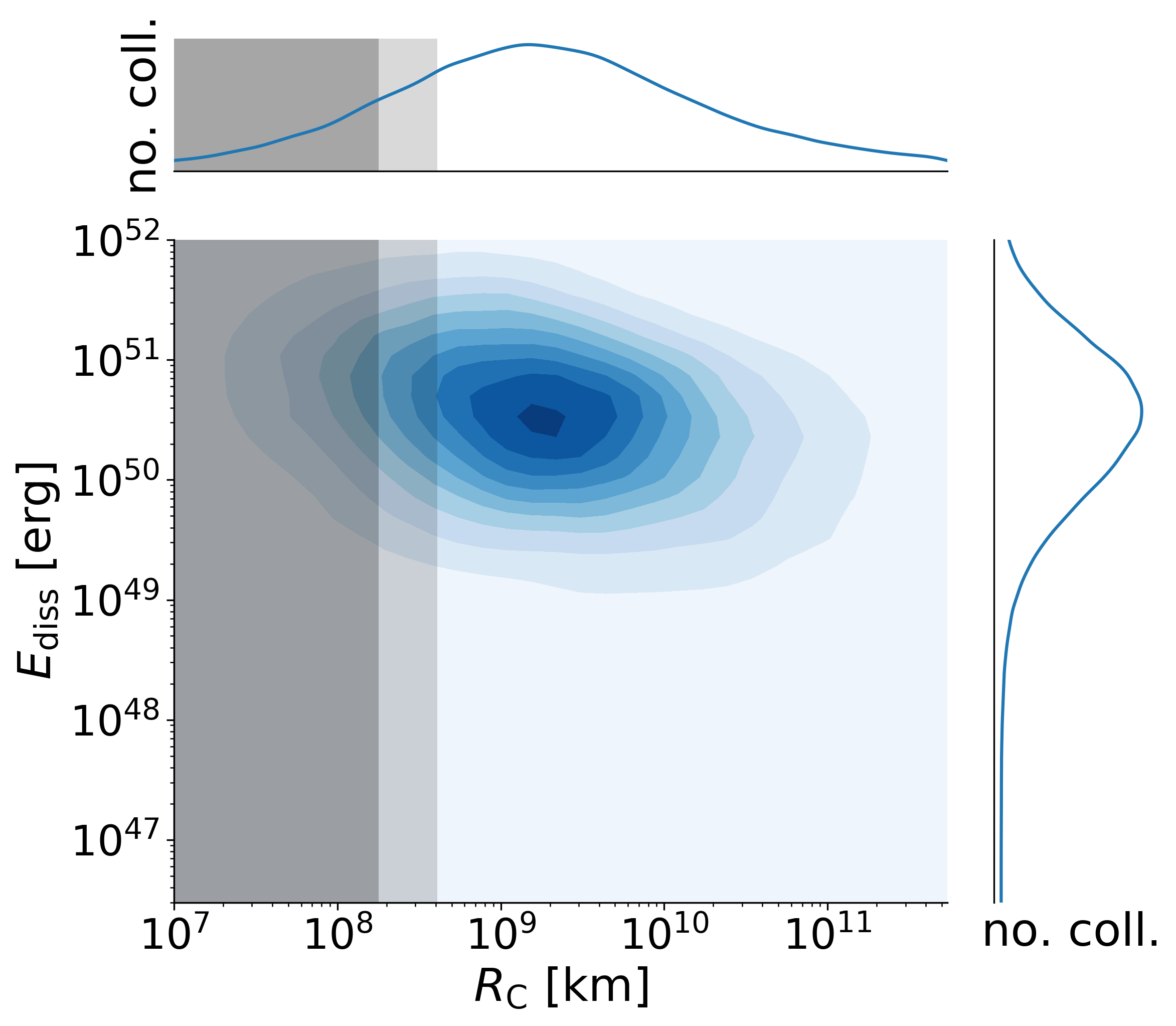}}
		\subfloat{\includegraphics[width=.35 \textwidth]{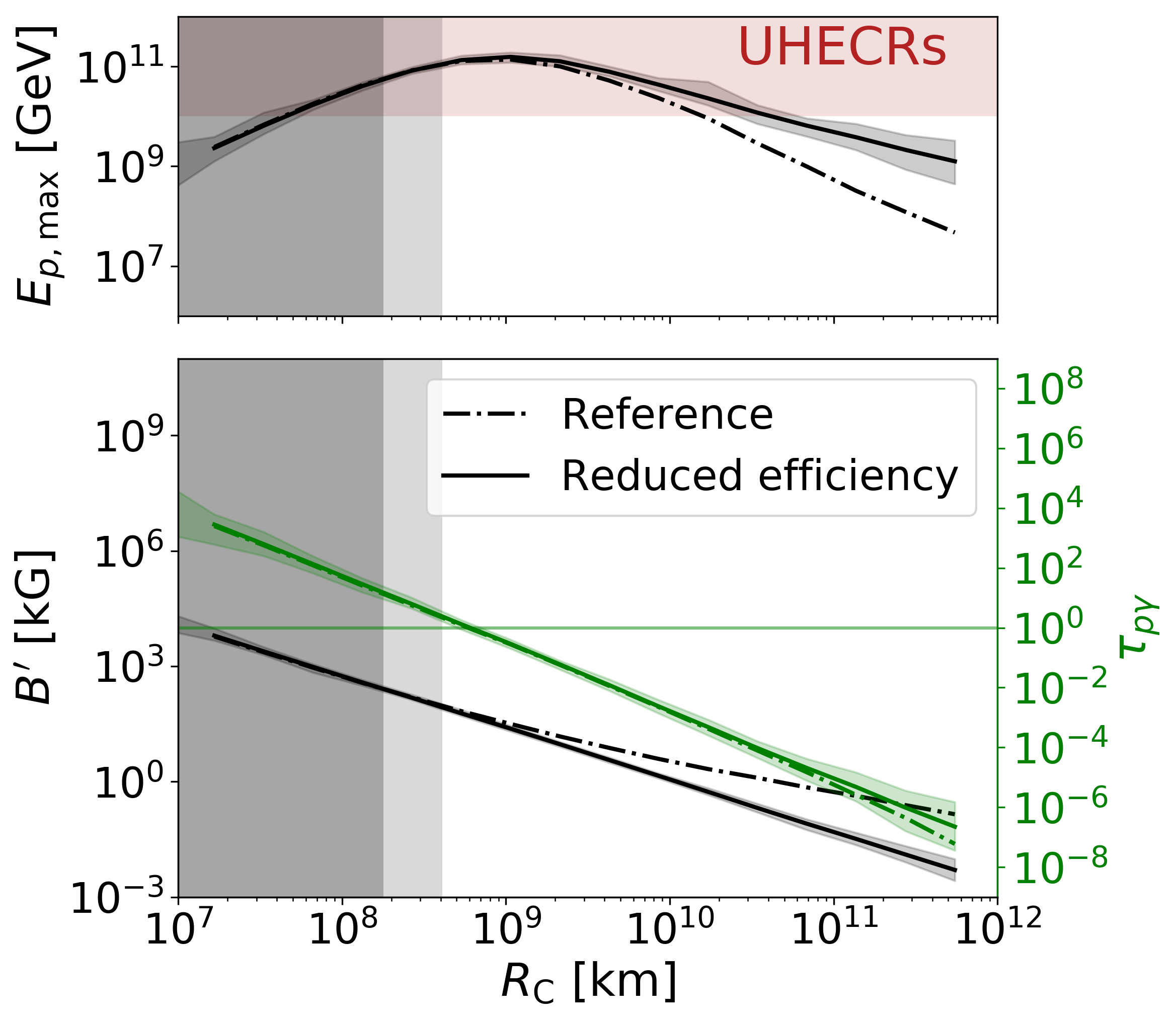}}
		
	}
	\hfill
	\makebox[\textwidth][c]{
		\subfloat{\includegraphics[width=.35 \textwidth]{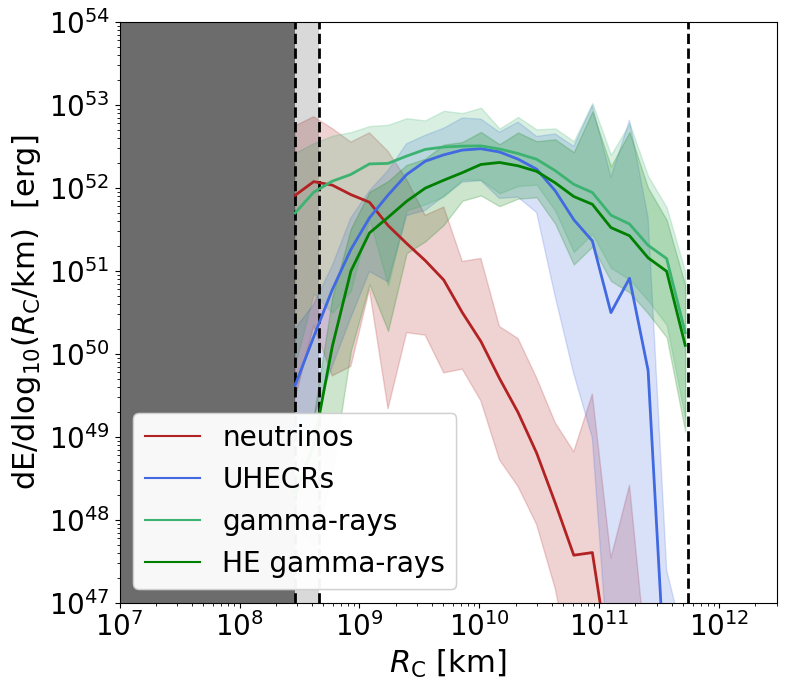}}
		\subfloat{\includegraphics[width=.35 \textwidth]{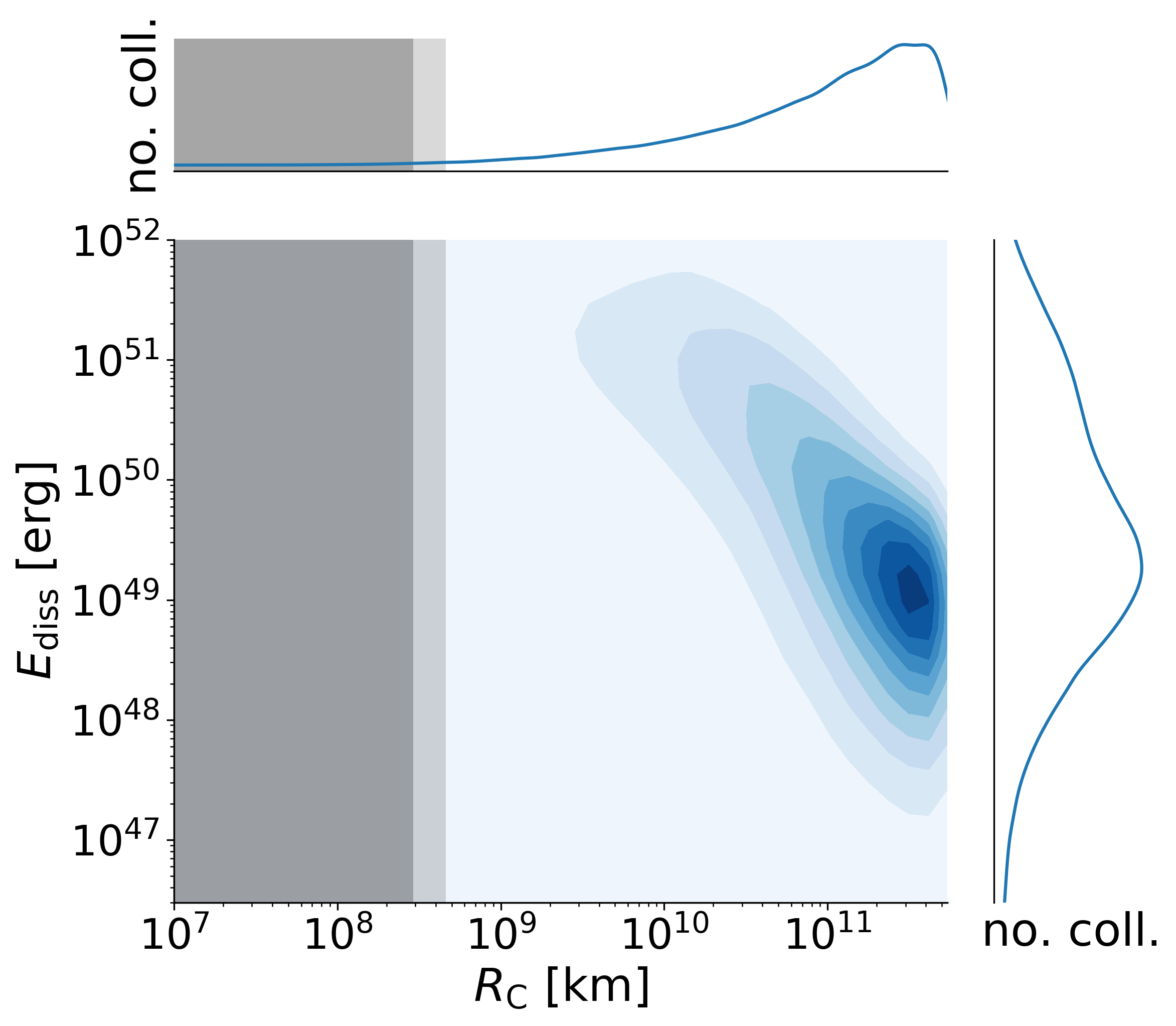}}
		\subfloat{\includegraphics[width=.35 \textwidth]{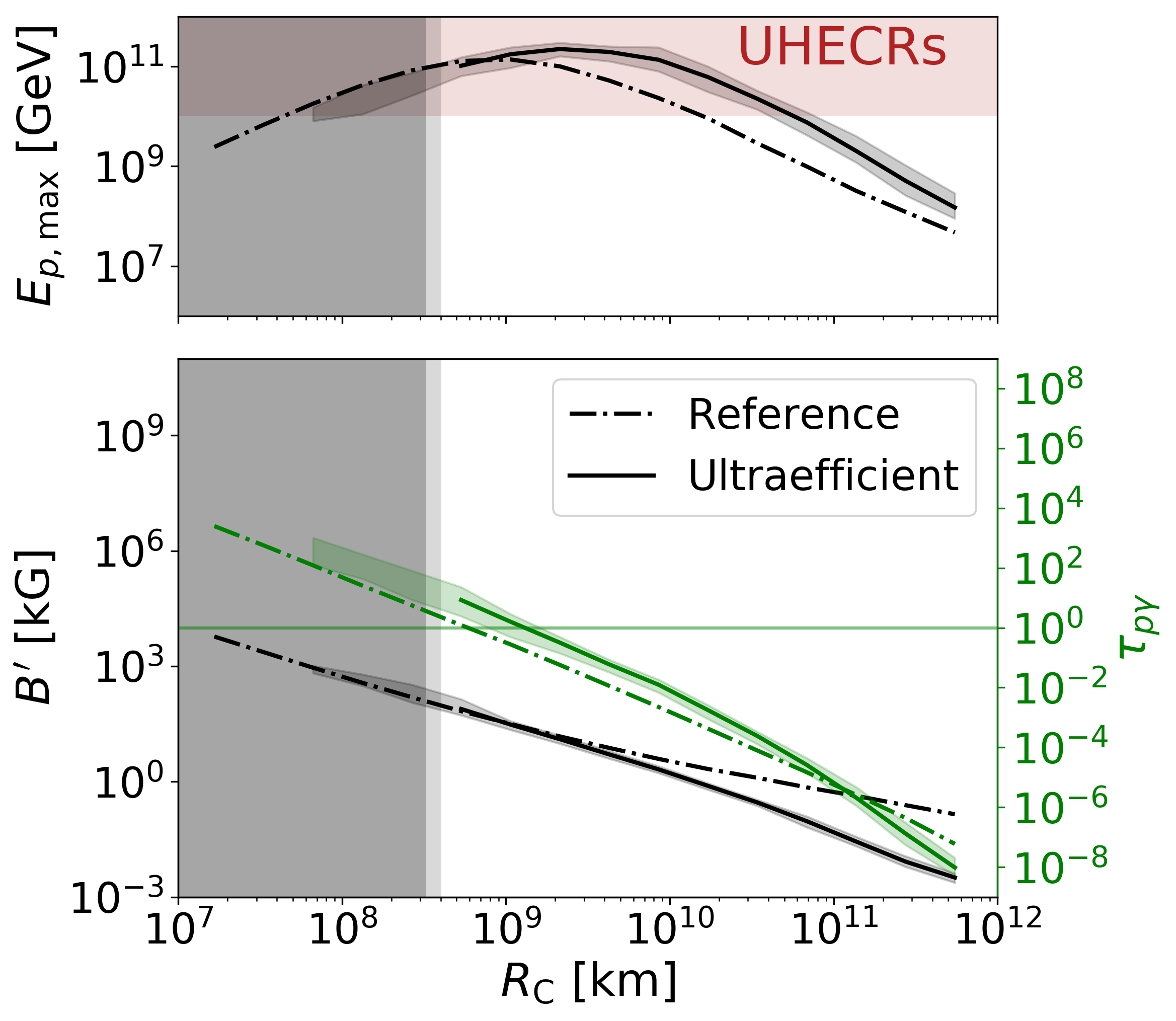}}
	}
	\hfill
	\makebox[\textwidth][c]{
		\subfloat[(a)]{\includegraphics[width=.35 \textwidth]{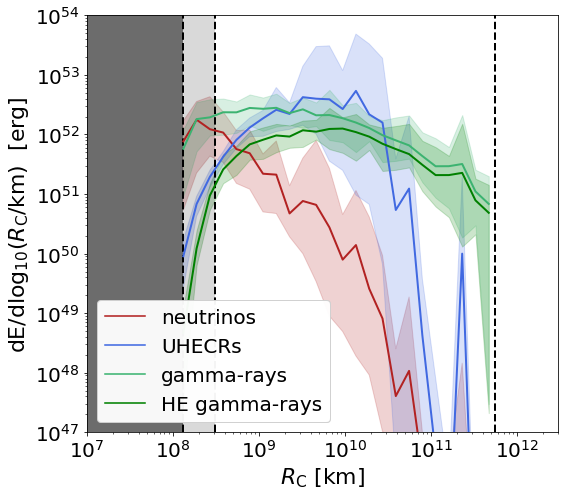}}
		\subfloat[(b)]{\includegraphics[width=.35 \textwidth]{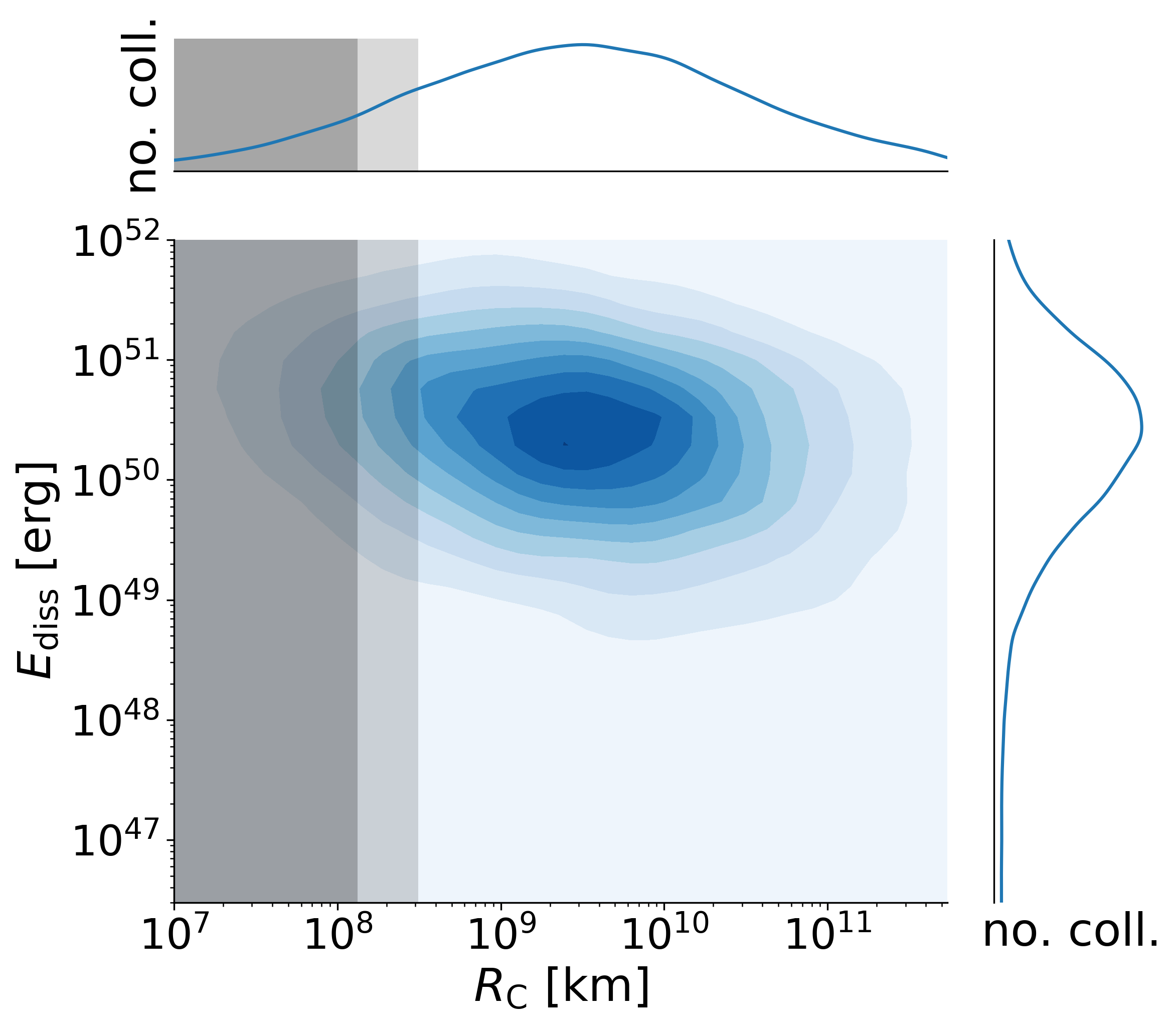}}
		\subfloat[(c)]{\includegraphics[width=.35 \textwidth]{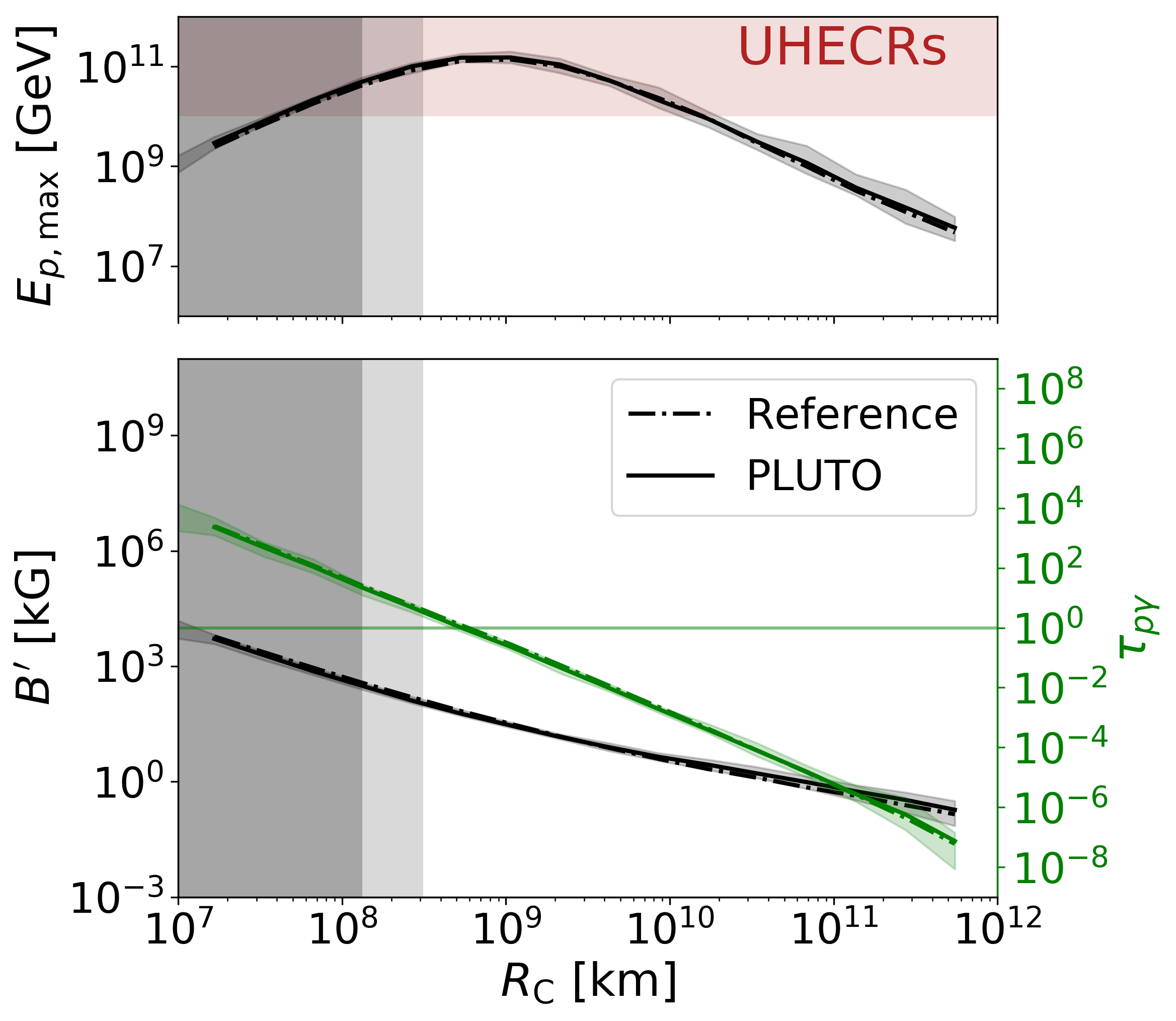}}
		
	}
	\caption{\label{fig:panel} Same format as \figu{panel_ref} but for the alternative models (in rows).  Note that the ensemble results for the {\em PLUTO} model (average and shaded areas lower row) have been only computed for twenty realizations due to computational restrictions.}
\end{figure*}

\subsection{Reduced Efficiency model}

The {\em Reduced Efficiency} model closely follows the {\em Reference} model from \sect{Reference_model} except that in each collision only a fraction $\eta = 0.5$ of the internal energy is dissipated as radiation. In the other models $\eta$ necessarily has to be smaller than one, as some energy is reconverted to kinetic energy.
We therefore chose this model to quantify the impact of the reduced efficiency $\eta$ compared to the {\em Reference} model ($\eta =1$).
During the collision the merged shell gets compressed according to \equ{shell_compression}; we assume that the emission takes place at this time. The remaining internal energy results in thermal expansion of the shell to the width $l_m = l_r + l_s$, which is the largest possible value which will never overlap with neighboring shells. The reduction of $\eta$ translates into a reduced overall efficiency $\epsilon \simeq 18\%$, see \tabl{model_results}.
As we choose to normalize the total gamma-ray output to the same value for each model, the inital kinetic energy budget has to be increased. As a consequence, the shells have higher density, and the photosphere moves to somewhat larger radii, see \figu{panel}, first row. 

In comparison to the {\em Reference} model, differences are visible mainly at large collision radii, \cf, solid versus dashed curves in \figu{panel}. We observe lower magnetic fields, slightly decreased optical depths, higher maximum proton energies and a generally more energy released in UHECRs. This effect comes from the thermal expansion of the merged shells that gain about a factor of two in width per collision, whereas in the {\em Reference} model the merged shell is still compressed. Consequently, the radiation density decreases at later stages of the fireball evolution, which is populated by previously collided shells. Since the maximal proton energies are higher and the photohadronic losses lower, the energy dissipated in UHECRs increases.  

\subsection{Ultra Efficient model}

In the Ultra Efficient Shock scenario proposed by \citet{Kobayashi:2001iq} the fraction $1-\eta$ of $E_\text{int}$ is reconverted to kinetic energy, which causes the merged shell to split into two shells. The dissipated energy $E_{\mathrm{int}}$ is calculated from \equ{ediss_multicoll} with $\eta=0.5$, the value same as in the {\em Reduced Efficiency} model. The Lorentz factors of the reflected shells are given by
\begin{align}
\begin{split}
 	\bar{\Gamma}_{\mathrm{r}} & = (M^2 + (m_{\mathrm{r}})^2 - (m_{\mathrm{s}})^2)/2m_{\mathrm{r}} M  \, ,\\
 	\bar{\Gamma}_{\mathrm{s}} & = (M^2 - (m_{\mathrm{s}})^2 + (m_{\mathrm{r}})^2)/2m_{\mathrm{s}} M  
 \end{split}
 \end{align}
in the center-of-mass frame, with $M = (m_{\mathrm{r}} \Gamma_{\mathrm{r}} + m_{\mathrm{s}} \Gamma_{\mathrm{s}} - \eta \cdot E_{\mathrm{int}} /c^2)/ \gamma_{\mathrm{m}}$. Converting back to the engine frame, this leads to
\begin{align}
\begin{split}
	\Gamma_{\mathrm{r}} & = \bar{\Gamma}_{\mathrm{r}} \gamma_{\mathrm{m}} -\sqrt{(\bar{\Gamma}_{\mathrm{r}}^2-1)(\gamma_{\mathrm{m}}^2-1)} \, \text{and} \\
	\Gamma_{\mathrm{s}} & = \bar{\Gamma}_{\mathrm{s}} \gamma_{\mathrm{m}} -\sqrt{(\bar{\Gamma}_{\mathrm{s}}^2-1)(\gamma_{\mathrm{m}}^2-1)} \, .
\end{split}
\end{align}
As in the {\em Reduced Efficiency} model, the particle acceleration and emission is assumed to take place in the compressed merged shell, calculated as in \sect{Reference_model}. After the collision, both shells are assumed recover their initial masses and widths, see \citet{Kobayashi:2001iq} for a more refined discussion of these assumptions.

In the {\em Ultra Efficient} model, shells repeatedly bounce off each other causing the fireball to thermalize and leading to an efficient conversion $\epsilon \simeq 36\%$ of the kinetic energy into radiation. For the same moderate dissipation per collision $\eta = 0.5$ this model yields twice the overall efficiency of the {\em Reduced Efficiency} model that reaches $\epsilon \simeq 18\%$. The consequent reduction of the engine's power requirements leads to a smaller deposition of kinetic energy in the afterglow. 

Due to the distribution of fewer initial shells across the same system size (to match the total burst duration) the widths and, more importantly, the separations are increased. Compared to the {\em Reduced Efficiency} model the shells travel longer before collisions appear and thus the bulk of collisions moves out to larger radii. This qualitative change of collision patterns and of average collision radii is visible in the middle panel of the second row in \figu{panel} (compared to the upper middle panel). Because the fireball thermalizes, the relative Lorentz factor difference of the colliding shells shrinks, leading to less dissipated energy per collision at late times of the evolution. Therefore, the collisions at large radii insignificantly contribute to the fireball's total emission (left panel of second row). The consequence is a reduced neutrino flux (see \tabl{model_results}) since neutrinos are predominantly produced in collisions close to the photosphere, where the densities and p$\gamma$ interaction rates are high (see also discussion in next section).

\newpage

\subsection{PLUTO model}

In the {\em PLUTO} model, each collision is individually simulated with {\sc PLUTO}, using $\eta =0.5$ and $d=0.7$ (see \sect{hydrodynamics} and \appr{PLUTO_methods} for more details). We analyze the resulting mass density profile to obtain the post-collision shell masses, widths and Lorentz factors by fitting box distributions to the obtained mass density profile and calculating the weighted average of the Lorentz factor profile.
In contrast to the {\em Reduced Efficiency} model, the thermal expansion of the shell(s) after the collision is not taken into account. 

The result for the {\em PLUTO} model is demonstrated in the third row of \figu{panel}. Note that the ensemble size has been reduced to twenty due to computational complexity. The distribution of collisions resembles the {\em Reference} model very closely. Only a small fraction (about $8\%$) of the collisions results in two post-collision shells that move the result closer to the {\em Ultra Efficient} model. The magnetic field strength, the optical depth to p$\gamma$ reactions, and the maximum proton energy are closer to the {\em Reference} model than the {\em Reduced Efficiency} model, since shells do not thermally expand after the collision. The overall dissipation efficiency is higher (21\% compared to 18\%) than in the Reduced Efficiency model, see \tabl{model_results}.

Despite choosing the initial conditions to enhance the occurrence of two-shell configurations as in the the {\em Ultra Efficient} model, our results indicate that these conditions can only be maintained under special circumstances.

\section{Impact on observations}
\label{sec:multimessenger}

So far, we have discussed and compared the evolution of the system, including the production of the astrophysical messengers, in the engine frame. Here we we focus on two observables: the expected quasi-diffuse neutrino flux in comparison with the current stacking limit, and the light curves. If not noted otherwise, we use $z=2$ in this section, and all quantities are given in the observer's frame.

\subsection{Neutrino flux prediction}

\begin{figure}[tb]
	\centering
	\includegraphics[width = \columnwidth]{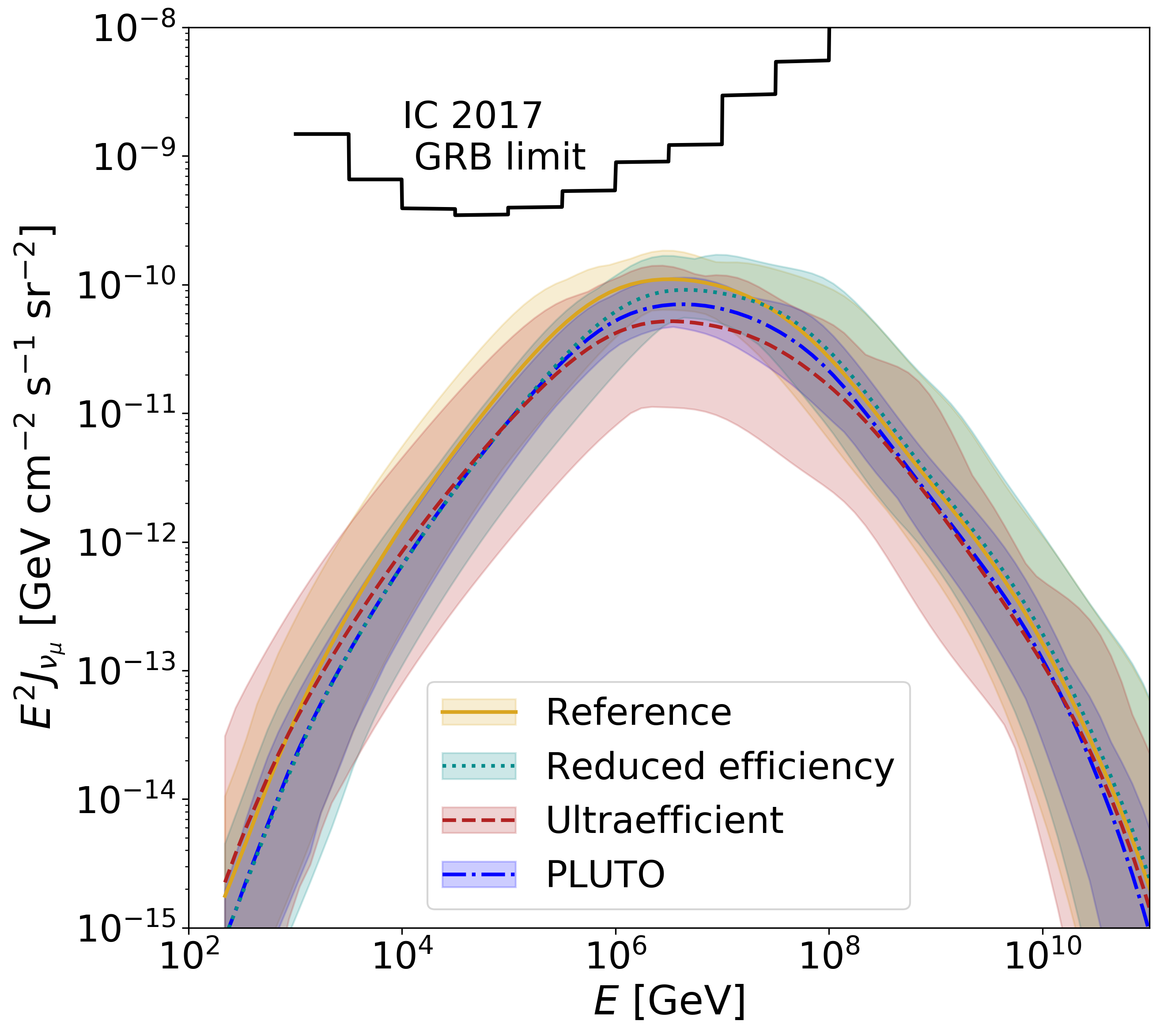}
	\caption{Model comparison of the expected neutrino fluxes excluding sub-photospheric contributions. We derive the all-sky quasi-diffuse $\nu_\mu + \bar{\nu}_\mu$ flux $J_{\nu_\mu}$ by scaling the fluence of one GRB $F_{\nu_\mu}$ as $J_{\nu_\mu} = (4 \pi)^{-1} \cdot F_{\nu_\mu} \cdot \dot{N}$ ,  assuming a rate of $\dot{N}=667$ identical GRBs per year at $z=2$ contributing to this flux (see \citet{Abbasi:2012zw,Aartsen:2017wea}). Curves correspond to the average, the shaded area to the range of neutrino fluxes obtained in the simulations. The GRB stacking limit, shown for comparison, is taken from the best limit obtained in \citet{Aartsen:2017wea}. \label{fig:nuflux}}
\end{figure}

A model comparison of the expected quasi-diffuse neutrino fluxes among the different models, including the ranges expected from the ensemble fluctuations, is shown in \figu{nuflux}. All predictions are relatively similar, except for the {\em Ultra Efficient} model, which has a substantial amount of collisions at larger radii, as previously discussed and summarized in Table~\ref{tab:model_results}. 

The expected neutrino flux is at the level of $10^{-11}$ to $10^{-10} \, \mathrm{GeV \, cm^{-2} \, s^{-1} \, sr^{-1}}$ (per flavor). Note the predicted flux in the equal-mass reference setup is about a factor of two higher than in the equal-energy setup used in \citet{Bustamante:2016wpu}, see discussion in \App~\ref{app:cases} and \figu{nufluxadd}. 
Since the GRB stacking search is basically background-free due to a probability density function including direction, time window and energy, we expect an improvement of a factor of seven to ten in the proposed next generation experiment IceCube-Gen2~\citep{Aartsen:2014njl}. Therefore, if taking the prediction at face value there is a good chance to find GRB neutrinos with the next generation detector.

There are several caveats here: the normalization method has been chosen to be compatible with the IceCube method to compute the stacking limit, namely: The fluence of one GRB is converted into a quasi-diffuse flux with a certain number of GRBs contributing per year (that can be related to the local GRB rate if the cosmological distribution of GRBs is known or measured), the baryonic loading of ten is an ad-hoc choice here, and the cosmological distribution of sources is not taken into account (in consistency with the stacking method, because for most GRBs the redshift is not known). The normalization of the flux prediction is therefore somewhat arbitrary, where some factors (number of GRBs per year and redshift) have been chosen in consistency with the stacking method and are expected to influence both the stacking limit and flux prediction in a similar way. 

The baryonic loading, however, is chosen in consistency with early estimates of the ultra-high energy cosmic ray (UHECR) energy budget, see \eg\ \citet{Waxman:1997ti}, assuming that GRBs are the sources of the UHECRs. Details, however, depend on the UHECR composition (especially in the light of recent Auger measurements, see \eg\ \citet{Aab:2016zth}), spectral shape, and the cosmic ray escape mechanism from the source. Recent approaches therefore fit the cosmic ray data obtaining the baryonic loading from a fit, see \citet{Baerwald:2014zga} and \citet{Biehl:2017zlw} in the one-zone model for protons and nuclei, respectively, and \citet{Boncioli:2018lrv} for low luminosity GRBs. A similar approach (without explicit derivation of the baryonic loading and parameter space scan) has been used for a multi-collision model in \citet{Globus:2014fka}. 

When comparing the energy output of $2$ to $4 \, \cdot 10^{52} \, \mathrm{erg}$ in UHECRs per GRB (\Tab~\ref{tab:model_results}, depending on the collision model) with the $3$ to $11 \, \cdot 10^{53} \, \mathrm{erg}$ to power UHECRs (see in Tab.~1 in \citet{Baerwald:2014zga}, depending on the source evolution), the baryonic loading (times gamma-ray luminosity) in our model will need to be a factor of $7-50$ higher if UHECR data are to be fit. This implies that the neutrino limit may be more severe than shown here, and that the collision model and escape mechanism in internal shocks can be already tested.  
A more detailed parameter space study is beyond the scope of this work and will be presented elsewhere.

If, on the other hand, GRBs are {\em not} the dominant source of UHECRs, the GRB neutrino stacking searches allow to derive an upper bound for the baryonic loading of GRBs and their contribution to the UHECR flux. Independent information can come from the spectral energy distribution (SED), which may be altered in the presence of hadronic processes (for instance in the Fermi-LAT band \citet{Asano:2011cp, Wang:2018xkp}). It is an open question how large baryonic loadings can be tolerated in self-consistent SED computations of GRBs (see \eg\ \citet{Asano:2007my,Asano:2008tc,Asano:2014nba}).

\subsection{Light curves}
\label{sec:light_curves}

\begin{figure}[tb]
	\centering
	\subfloat[(a)]{\includegraphics[width=\columnwidth]{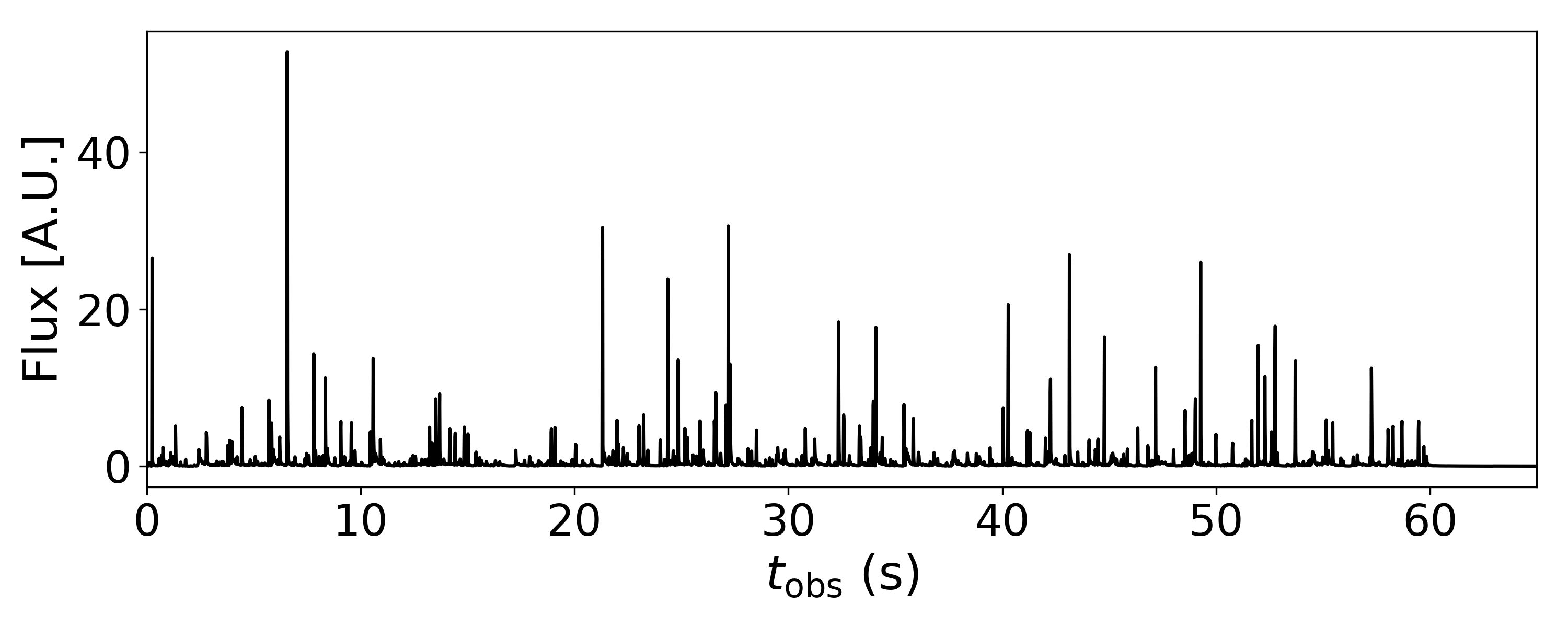}}\\
	\subfloat[(b)]{\includegraphics[width=\columnwidth]{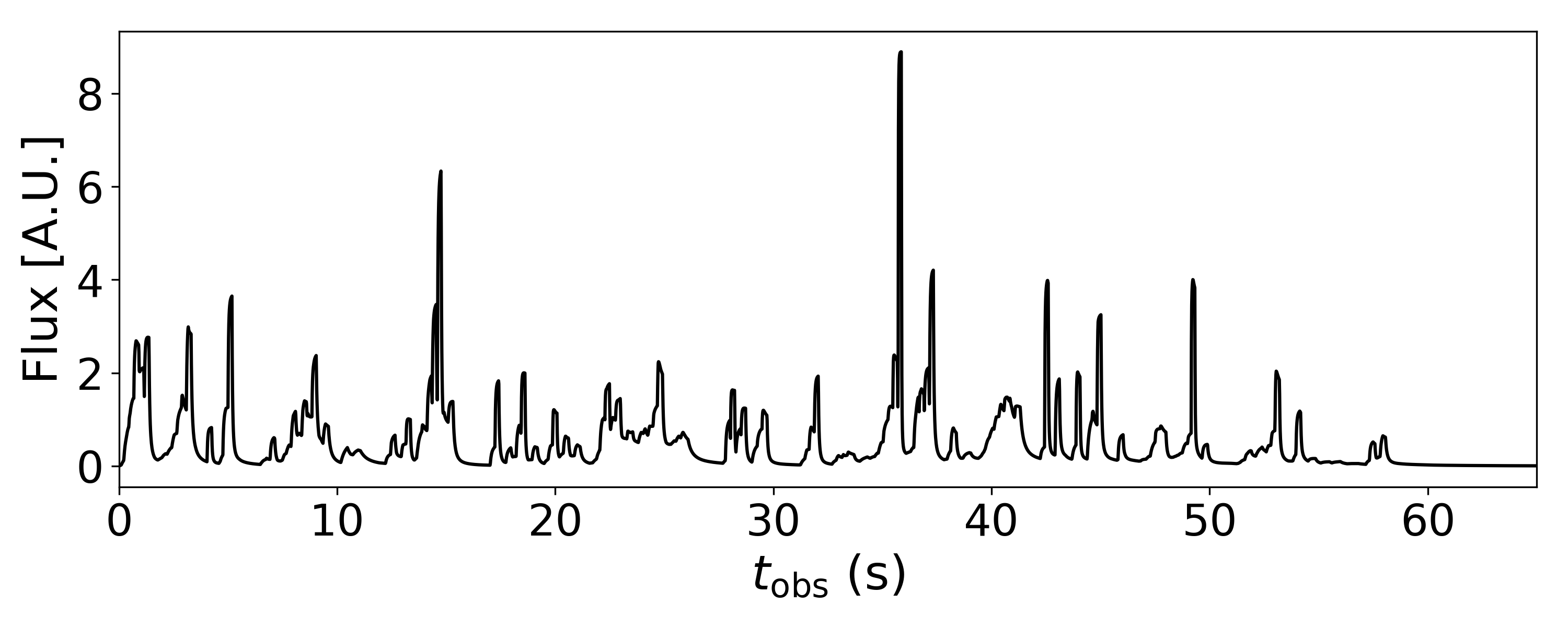}}
	\caption{Energy flux as a function of time for one example GRB for (a) the Reference model, (b) the Ultra Efficient model.}
	\label{fig:lightcurves}
\end{figure}

Another interesting question concerns the impact of the collision model on GRB light curves that are predicted by the multi-collision models. \figu{lightcurves} shows two examples generated with the {\em Reference} model and the {\em Ultra Efficient} model; the {\em Reduced Efficiency} and {\em PLUTO} model light curves are very similar to the {\em Reference} model. Recall that the total burst durations are chosen to be equal by construction. We also checked different wavelength bands (not shown here), but did not observe significant differences (such as time delays) because of the stochasticity of the setup chosen here -- in consistency with \citet{Bustamante:2016wpu} (see discussion therein). 

The light curves for the {\em Reference} model are composed of many numerous thin peaks of comparable height, similar to \citet{Kobayashi:1997jk}. In contrast the {\em Ultra Efficient} model produces light curves that are dominated by pronounced, broad pulses -- modulated by some time variability, see also \citet{Kobayashi:2001iq}. The widening of the pulses originates from the wider initial shells, and hence comes from the properties of the engine. A few collisions with high energy dissipation result in a few pronounced pulses contributing most of the photon counts. 

Despite qualitative differences in the shape of the light curves, the time variability, derived by dividing the total burst duration by the number of collisions, is similar for both models (see \tabl{model_results}). 

\section{Discussion}

\subsection{Dissipation efficiency}
\label{sec:eff}
 
\revise{
Originally, the {\em Ultra Efficient} collision model \citep{Kobayashi:2001iq} was introduced to increase the emission efficiency of the prompt fireball phase. This is achieved at the cost of reduced efficiency for individual collisions and a higher count of multiple collisions within the fireball. We indeed find a higher efficiency, see \Sec~\ref{sec:hydrodynamics} and \Tab\ref{tab:modelcomp} but only for a small subset of fine-tuned shell parameters. The more realistic {\em Pluto} and the {\em Reduced Efficiency} models result in lower efficiencies and as a consequence require the initial kinetic energy to be significantly higher for the same gamma-ray output. The {\em Reference} model reaches efficiency values comparable to the {\em Ultra Efficient} model due to high single-collision efficiencies under the ideal assumption of $\eta=1$. The efficiencies of either model are higher than the 1--15 \% discussed, for example, in \citet{Peer:2015eek} and in the multi-messenger context in \citet{Globus:2014fka}. This is related to the combination of a large spread in initial shell Lorentz factors and a equal-mass outflow, two features that are known to increase the efficiency \citep{Sari:1997kn,Kino:2004uf}. Those are technical parameters of the internal shock fireball model that, at present, lack clear experimental constraints.}

\revise{ 
The inferred efficiency of detected GRBs is usually based on observations of the energy content in afterglows. Comparing the inferred efficiency to that of the internal shock model has given rise to criticism. This statement is, however, derived from an incomplete observation of the afterglow spectrum and recent observations of high-energy emission \citep{Acciari:2019dxz, Arakawa:2019cfc} support the arguments \citep{Fan:2006sx, Beniamini:2015eaa, Beniamini:2016hzc} that these estimates may underestimate the energy content in the afterglow. }

\subsection{Model assumptions}

\revise{We discuss our choices of model parameters in the context of previous studies. The choice of 1000 initial shells (leading to approximately 1000 collisions) is at the upper end of the typical range and may lead to a short time-variability. 
The number of collisions (and thus the number of initial shells) and the short time variability are directly related by $N_{coll} \approx  T_{90} / t_{\mathrm{min, var}}$. When comparing to observations, we find that \citet{MacLachlan:2012cd} show values ranging from $T_{90} / t_{\mathrm{min, var}} \approx  100$ to $1000$. In \cite{Golkhou:2014zta, Golkhou:2015lsa} the values seem to be more in the range of $10 - 100$, with a significant fraction of values smaller than that. However, this critically depends on $t_{\mathrm{min, var}} $ -- and in practice, the observable short-time variability is constrained by the instrument's response, limiting the possibilities to rule out very small values.
On the same note,there is currently no evidence for a very high initial spread of shell Lorentz factors (as assumed in our log-normal distribution). However this choice is not uncommon in attempts to reach high dissipation efficiencies~\citep{Kobayashi:1997jk, Kobayashi:2001iq, Bustamante:2016wpu}. Since in this paper we study the impact of modifications to the single-collision model on the multi-messenger production and emission, we do not attempt to depart from these previous assumptions for the sake of comparability with previous results in \citet{Bustamante:2016wpu}.}

\revise{Another feature that can be compared to other GRB-related studies is the required engine kinetic energy. For our assumptions the (isotropic-equivalent) engine power of a few times $10^{54}$~ergs is required, which seems to be at the upper end of observations \citep{Gruber:2014iza}. As outlined in \citep{Beniamini:2015eaa}, the actual kinetic energies may be higher but it is disputed if the timescale applied in \citet{Beniamini:2015eaa} is appropriate.
The recent discovery of inverse Compton emission from GRB 190114C also yields values in this ballpark~\citep{Acciari:2019dbx}. In general, such energies are unavoidable to power the UHECR flux, see \eg{} Sec.~2 in \citet{Baerwald:2014zga}. At present, there is no direct observational evidence for cosmic ray emission from Gamma-Ray Bursts, due to the lack of high-energy neutrino observations in coincidence with detected GRBs -- although in our model neutrinos are not expected for the current generation of neutrino detectors, see \figu{nuflux}.}

\revise{In summary, some of our assumptions seem extreme, especially in their combination. This has been partially motivated by the comparison to the existing literature, partially by the paradigm that GRBs would be the dominant source of UHECRs. }

\section{Summary and conclusions}
\label{sec:summary}

We have studied the production of multiple astrophysical messengers (gamma-rays, neutrinos, and UHECR protons) in the Gamma-ray Burst internal shock scenario. We have used a multi-collision scenario involving a set of plasma shells emitted from the central engine with different Lorentz factors, where we have computed the production/emission of the messengers in each collision individually. The focus of this work has been the impact of the collision model between two shells on the multi-messenger production in the entire fireball, where we have tested different scenarios:
\begin{description}
 \item[Reference] \ Plasma shells merge (inelastic collision) and all internal energy is dissipated in non-thermal radiation 
 \item[Reduced Efficiency] \ Plasma shells merge and a fraction of internal energy is dissipated (the rest goes into thermal expansion).
 \item[Ultra Efficient] \ Plasma shells do not merge and a part of the internal energy is reconverted into kinetic energy that separates the shells.
 \item[PLUTO] \ Fate of plasma shells determined for each collision individually from hydrodynamical simulations
\end{description}
We have also tested different assumptions for the behavior of the central engine (outflow at equal-mass or equal-energy rate), while fixing the observables of the GRB (gamma-ray luminosity, duration, time variability, observed broken power-law spectrum) for better comparability. Compared to earlier works, we have also studied the impact of ensemble fluctuations from the stochastic setup on the results. 

For all models, we have recovered the qualitative behavior of the multi-messenger emission, the different astrophysical messengers are emitted from different regions of the same object: neutrinos come from the innermost collision radii where the radiation densities are highest, gamma-rays come from a wide range of collision radii, where most energy is dissipated, HE gamma-rays prefer large collision radii because they can escape the optically thin shells, and UHECRs come from intermediate radii requiring balanced magnetic fields strengths that sustain acceleration up to high energies without introducing too strong synchrotron losses. Additional contribution to the neutrino flux may come from below the photosphere; given the lack of observational constraints on the photon density and the maximal proton energies, the current modeling framework does not permit to include this component. Thus, our neutrino flux estimations are conservative, lower limits in this regard. 

Substantial differences in the distribution of collision radii have been found in the Ultra Efficient case, coming with a quantitative impact on \eg\ the neutrino flux, UHECR production radius and shape of the GRB lightcurve. In that case, a smaller number of initial shells (with larger widths and inter-shell spacings) bounce frequently off each until the system thermalizes. Consequently, more collisions occur at large radii compared to the other models, while more energy is dissipated in the first collisions. For this case we have found the highest dissipation efficiency (kinetic energy converted into radiation) of about 40\% and about a factor of two lower neutrino flux.

We have scrutinized the assumptions of the Ultra Efficient scenario in hydrodynamical simulations with {\sc PLUTO}. We have found that the bounce back (instead of merging) of shells only hold for very specific conditions: A collision only results in two shells if energy dissipation is not included, the shells are set up with roughly equal masses and there is a large ratio between Lorentz factors of the rapid and the slow shell. These optimal conditions are occur less likely at later stages of the fireball evolution where multiple preceding collisions gradually thermalize the system. In that case most collisions result in a single dispersing shell. We have coupled the simulation of the fireball with the explicit {\sc PLUTO} simulation for the collisions of individual shells, and for this study included energy dissipation in the PLUTO simulations. That has demonstrated that the emergence probability of two-shell configurations -- as assumed in the \textit{Ultra Efficient} scenario -- is very rare ($\sim 8\%$), which means that our {\em Reference} model result is roughly recovered.
As a result, the promise of an increased efficiency (like in the {\em Ultra efficient} model) can't be kept and the efficiency-related problems of the fireball model remain unsolved. 

In all cases, the expected neutrino flux has been in the range $10^{-11}$ to $10^{-10} \, \mathrm{GeV \, cm^{-2} \, s^{-1} \, sr^{-1}}$ (per flavor) for a baryonic loading of ten, which is potentially in reach of the next generation neutrino observatories, such as IceCube-Gen2. However, we have also noted that the normalization of this flux is somewhat arbitrary. For example, if GRBs are to be the origin of the UHECRs, the baryonic loading will emerge as a deduced quantity from a fit to the observed UHECR flux and composition. From energy budget considerations, we anticipate that the UHECR output will be not sufficient to power UHECRs with the current parameters and the assumptions for the UHECR escape. Further studies will therefore be needed to establish when and if neutrinos can rule out the UHECR origin from GRBs in the framework of internal shock multi-collision models, and to what extent an observation will depend on the UHECR composition, the escape assumptions, and the collision model.

\acknowledgments
We are grateful to Maxim Barkov and Daniel Biehl for fruitful discussions. This work has been supported by the European Research Council (ERC) under the European Union’s Horizon 2020 research and innovation programme (Grant No. 646623). AF completed parts of this work as JSPS International Research Fellow.
\software{PLUTO v4.2 \citep{Mignone:2007iw}, NeuCosmA \citep{Biehl:2017zlw}}

\bibliographystyle{aasjournal}
\bibliography{references}

\begin{appendices}
Here we provide additional information referred to in the main text.

\section{PLUTO Methods}
\label{app:PLUTO_methods}

We simulate the collision with \textsc{PLUTO} v4.2 \citep{Mignone:2007iw}, an open source modular numerical code designed to describe high MACH number astrophysical flows containing discontinuities. 
Since we neglect magnetization (studies including magnetic fields have \eg\ been presented in \citet{Mimica:2006mc, Mimica:2009na}), we apply the Relativistic HydoDynamics (RHD) physics module and the ideal equation of state. The system is evolved in the rest frame of the contact discontinuity (CD, variables in the CD frame are denoted with the superscript $ ^\prime$), the gas adiabatic index is set to $\hat{\gamma} = 4/3$ (relativistic) if the Lorentz factor of the fast shell in the CD frame is above two ($\Gamma^\prime_{r} > 2$) and $\hat{\gamma} = 5/3$ (non-relativistic) else. The simulations are performed using Cartesian geometry, a linear reconstruction, Hancock time-stepping and the CD-restoring HLLC approximate Riemann solver (for more details about these methods, see the PLUTO user guide).

The simulation starts at $t^\prime = 0$, the time at which both shells get in contact. The coordinate system is defined such that the bondary between initial shells is at $x^\prime =0$ for $t^\prime = 0$. The fast shell ($x^\prime < 0$) has a positive velocity and the slow shell ($x^\prime > 0$) negative. The shells are surrounded by low density plasma ($\rho / \rho_{\text{shell}} = 10^{-12}$), moving at the same speed as the neighboring shell. The simulation is stopped after both the reverse and the forward shocks have crossed the shells.

To account for energy dissipation due to acceleration of cosmic rays and/ or electrons, we add a term to the differential equation solved by PLUTO. In a very simplified treatment, we assume the dissipation rate of internal energy at a given point in time and space to be proportional to the available internal energy. We withdraw energy from the system until a threshold $\eta^* E_{\mathrm{int, th}}$ is reached, setting the dissipation rate to zero afterwards: 
\begin{align}
	\frac{\partial \rho  \, e}{\partial t} = \begin{cases} 
	- \chi \cdot \rho \, e & E_\mathrm{diss, total} \leq \eta^* E_{\mathrm{int, th}} \\
	0 & \mathrm{else} \end{cases}
\label{equ:ediss_pluto}
\end{align}
where $\rho$ is the rest mass density, $e$ the internal energy per unit mass (thus $\rho e $ is the internal energy density) and $\chi$ a freely chosen parameter, $\eta^*$ controls the total dissipated energy by relating it to $E_{\mathrm{int, th}}$, the dissipated energy as predicted by the internal shock model (\equ{ediss_multicoll}). We set $\chi = 10\ \eta^*  / t_{\mathrm{shock}}^\prime$ and $\eta^* = \eta$ in the multicollision modeling. 

\section{Additional Figures: Hydrodynamic simulations}
\label{app:addfigs}
\begin{figure}
	\centering
	\includegraphics[width = .5\textwidth]{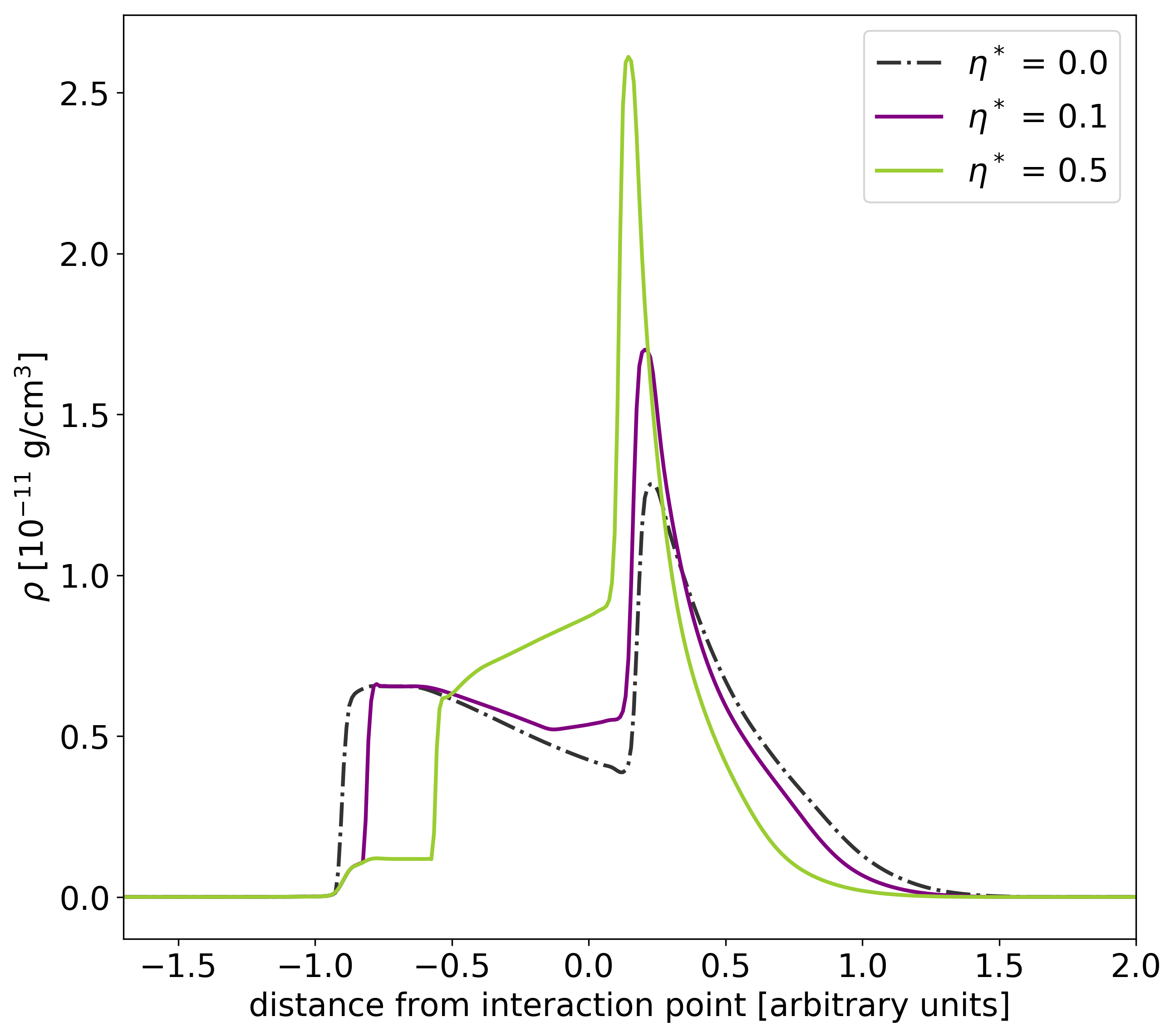}

	\caption{Snapshot of the mass density profile (in the frame of the contact discontinuity) of a single two-shell collision for three choices of $\eta^*$ in \equ{ediss_pluto}. In this example we set $\Gamma_{\mathrm{r}} /\Gamma_{\mathrm{s}} = 6.0 $ and $m_{\mathrm{r}} = m_{\mathrm{s}}$, the snapshot is taken at $t^\prime =  t^\prime_{\mathrm{shock}} = 2.8 \, \mathrm{s}$ (which equals the shock crossing time of the reverse shock). For $\eta^* = 0$, both the reverse and the forward shocks have crossed the respective shells. For higher $\eta^*$, the step in the mass density profile at the left side of the figure indicates, that the reverse shock has not yet completely crossed the slow shell, implying that energy dissipation slows down the shocks. Also, the dip between the two shells, which is clearly visible for $\eta^* = 0$, disappears for higher $\eta^*$. We conclude that energy dissipation reduces the separation between the shells after the collision. }
\label{fig:pluto_appendix_singlecoll}
\end{figure}

As an addition to \sect{hydrodynamics}, we illustrate the impact of energy dissipation on the evolution of the hydrodynamic system in \figu{pluto_appendix_singlecoll} and discuss the impact of the relative shell widths on the final shell configuration. 

\begin{figure}
	\centering
	\includegraphics[width = .5\textwidth]{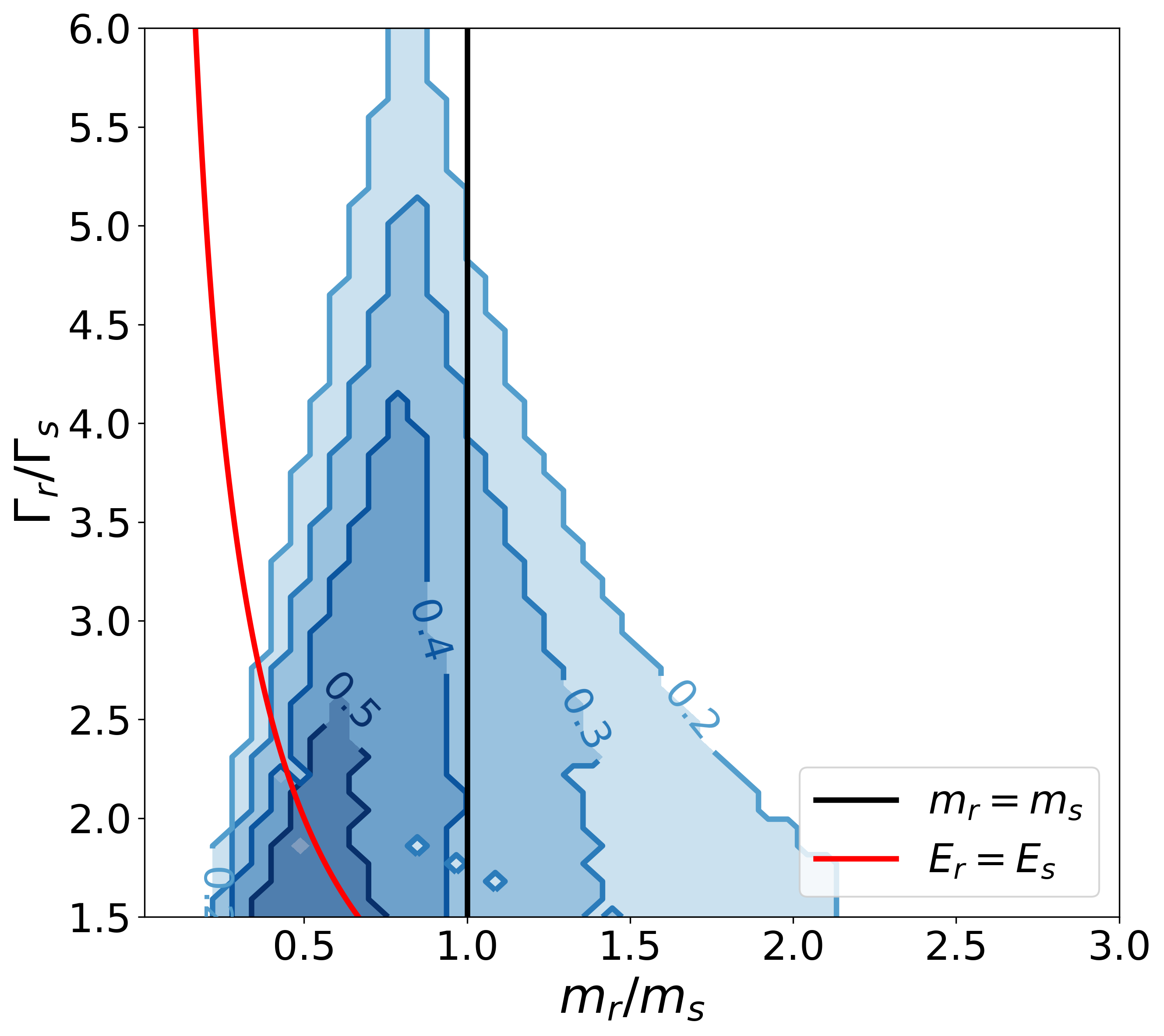}
	\caption{Same as \figu{parameterscan} without energy dissipation but with non-equal shell widths: $l_r = 0.1 \cdot l_s$ in the source frame. Again, the curves are height lines of the dip depth $d=(\rho_{\mathrm{max}} - \rho_{\mathrm{min}})  / \rho_{\mathrm{max}} $. In contrast to \figu{parameterscan}, we observe low ratios $\Gamma_r / \Gamma_s$ to increase $d$, also slightly smaller $m_r / m_s$ are favored.}
\label{fig:pluto_appendix_parameterscan}
\end{figure}
In \sect{hydrodynamics} we equal shell widths $l_r = l_s$ in the source frame. Since the relative shock and rarefaction wave timescales define the produced mass density profile \citep{Kino:2004uf}, it is natural to expect an impact of relative shell widths on the final shell configuration. The parameter scan for $l_r = 0.1 \cdot l_s$ is shown in \figu{pluto_appendix_parameterscan}. In contrast to the findings in \sect{hydrodynamics}, we observe low Lorentz factor ratios $\Gamma_r / \Gamma_s$ to increase the dip depth $d$. 
Without a detailed quantitave discussion, we want to explain this result in a qualitative way, looking at the changing ratios of $\rho_s / \rho_r$ and $l^\prime_s > l^\prime_r $ (primed quantities refer to the frame of the contact discontinuity (CD)). While for $l_r = l_s$ generally  $\rho_s > \rho_r$ and $l^\prime_s < l^\prime_r $, both relationships change to the opposite for $l_r = 0.1 \cdot l_s$. As a result, the dip separating the two post-collision shells is created in the slow shell instead of the fast one. While before increasing the ratio of the shock crossing times $t^\prime_{\mathrm{shock, }rs} / t^\prime_{\mathrm{shock, }fs}$ ($rs$ refers to the reverse shock, $fs$ to the forward shock) led to more pronounced dips in the mass density profile, this will now be the case for $t^\prime_{\mathrm{shock, }fs} / t^\prime_{\mathrm{shock, }rs}$. As a result, the opposite changes in $\Gamma_r / \Gamma_s$ are favorable for the formation of two, distinct shells and the plot flips with respect to the y-axis. 

This result illustrates the strong dependence of the resulting mass density profile on the individual shell parameters and the difficulty of finding universally applicable predictions. It supports our approach in the multi-collision modeling, where we account for the variety of possible impacts by modeling each collision individually with \textsc{PLUTO} instead of using simplified, universal formulas. 
 \newpage

\begin{figure*}
	\centering
	\makebox[\textwidth][c]{
		\subfloat{\includegraphics[width=.35 \textwidth]{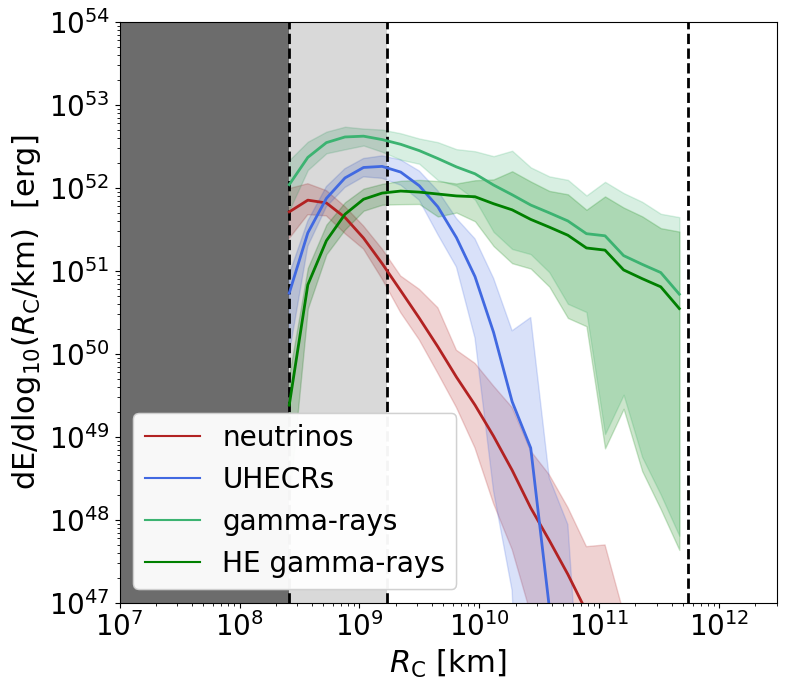}}
		\subfloat{\includegraphics[width=.35 \textwidth]{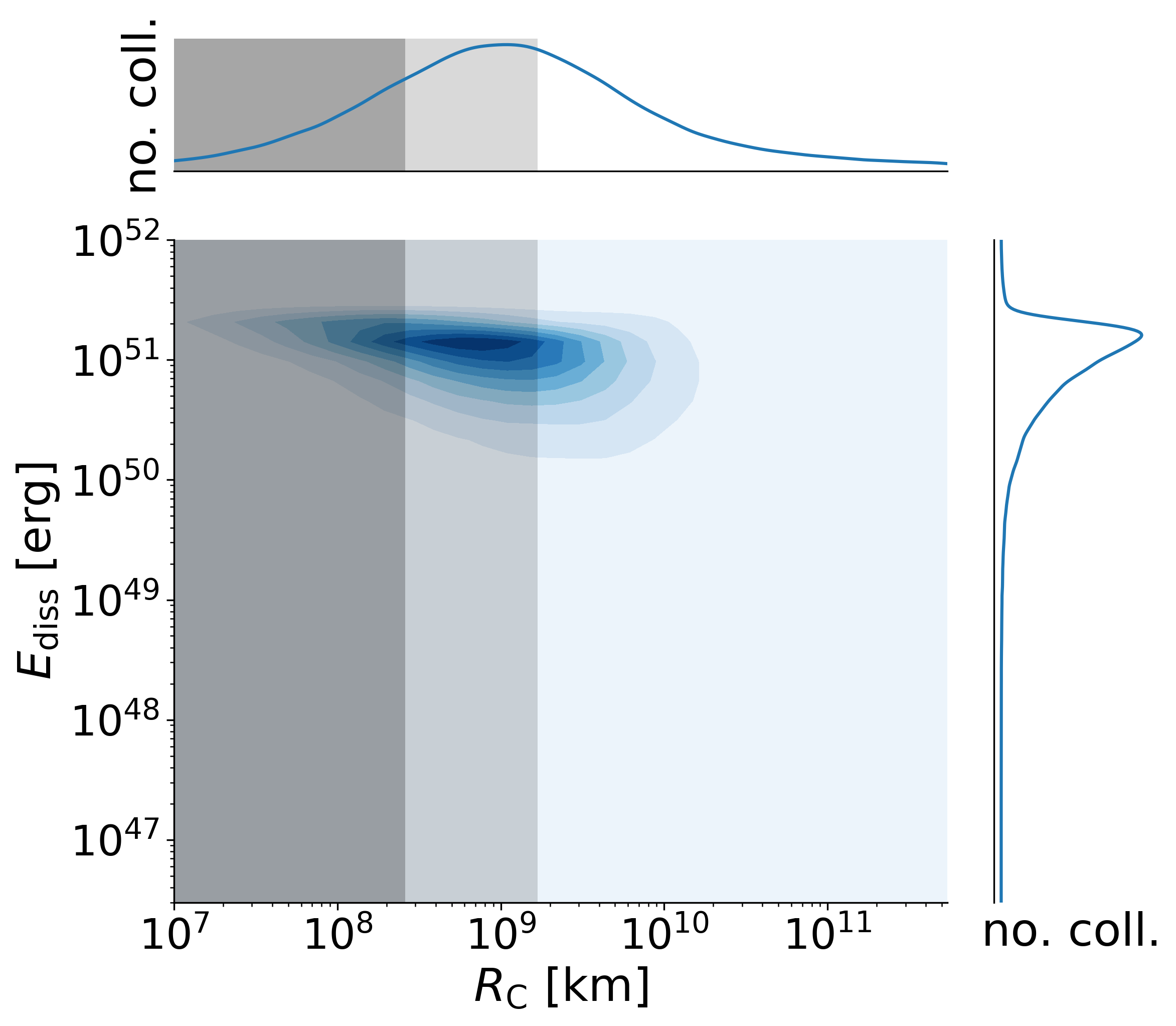}}
		\subfloat{\includegraphics[width=.35 \textwidth]{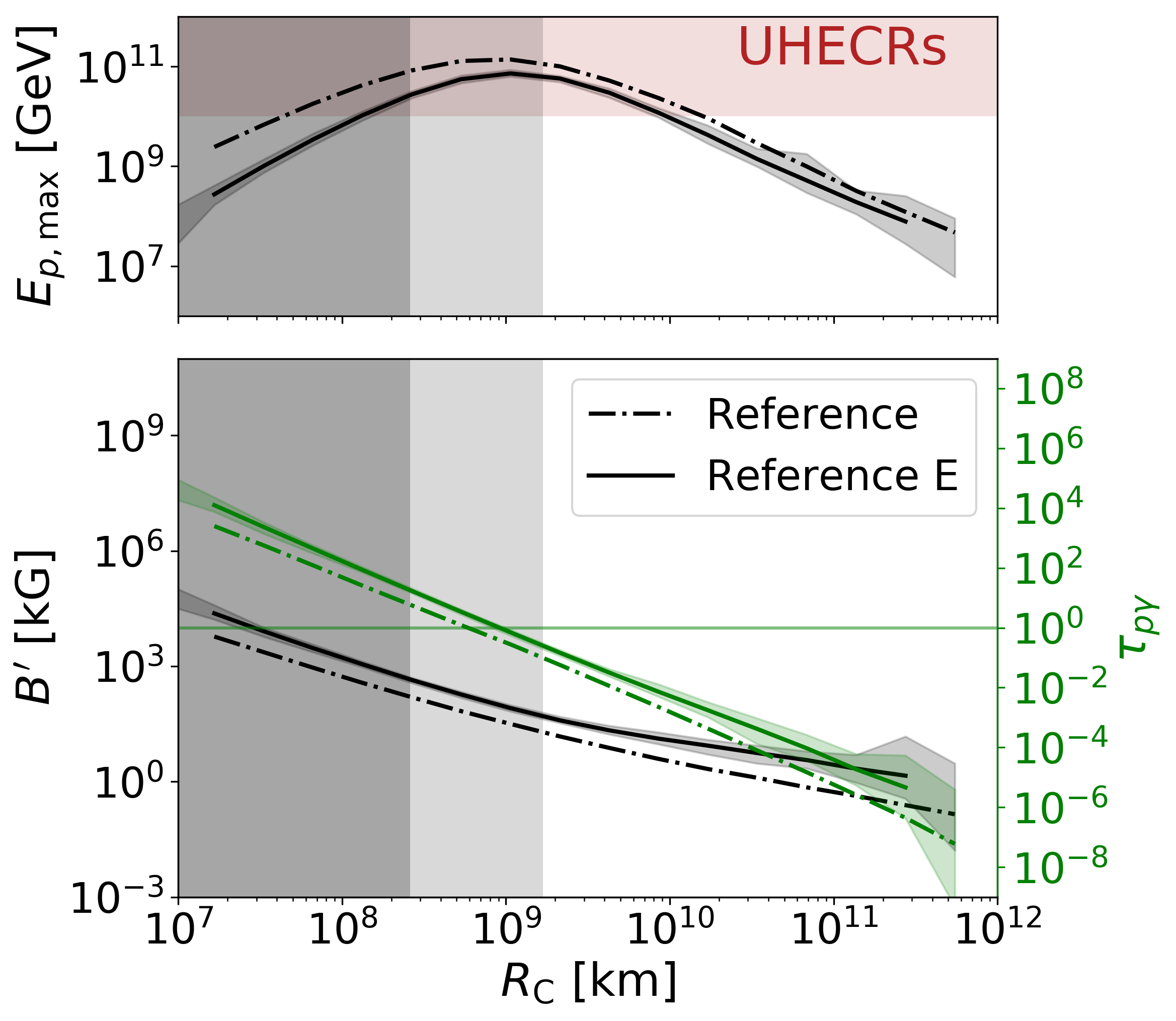}}
		
	}
	\hfill
	 
	\makebox[\textwidth][c]{
		\subfloat[]{\includegraphics[width=.35 \textwidth]{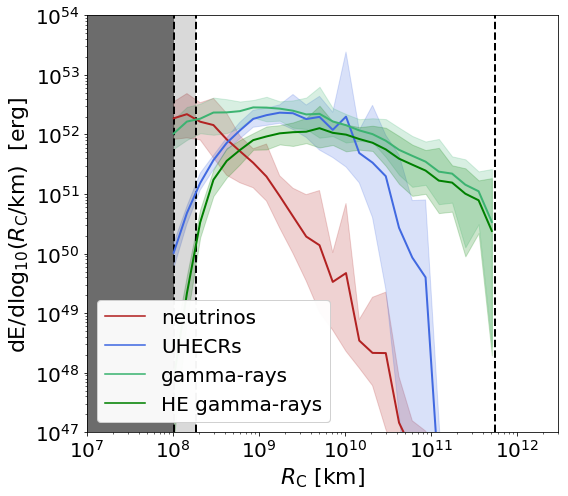}}
		\subfloat[]{\includegraphics[width=.35 \textwidth]{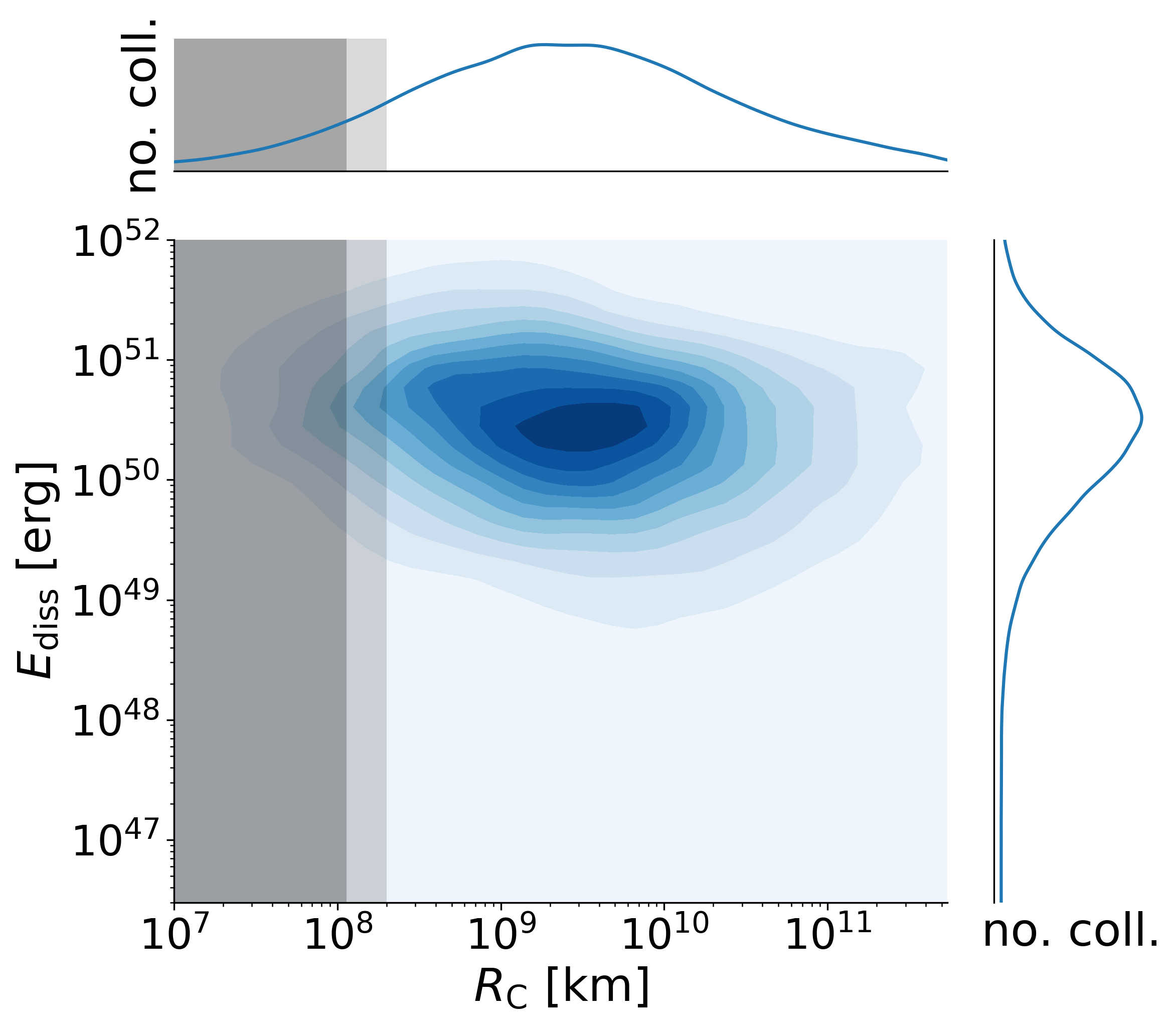}}
		\subfloat[]{\includegraphics[width=.35 \textwidth]{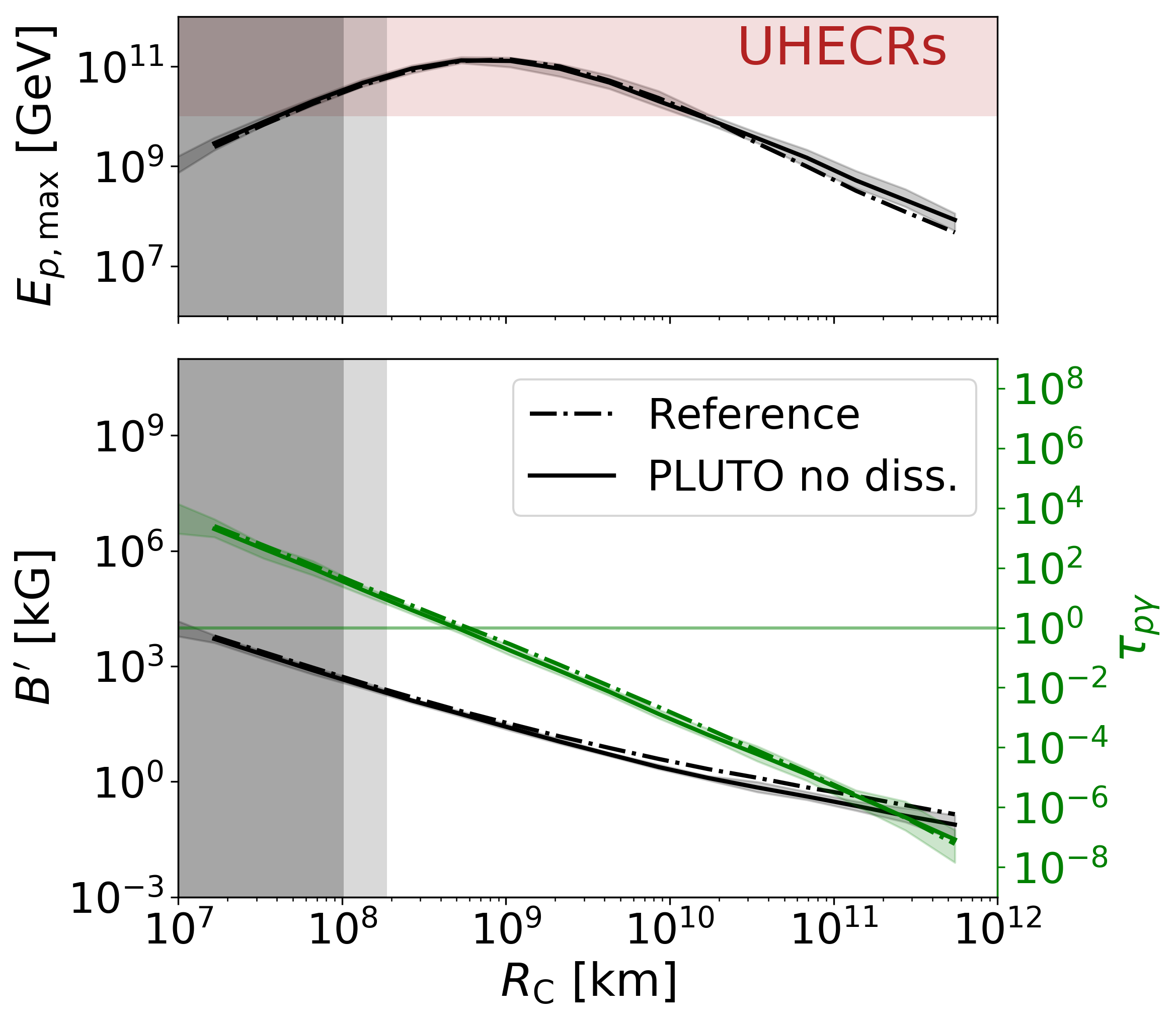}}
		
	}
	 
	\caption{\label{fig:panel_ref_E} Upper panel: {\em Reference} model, but with a source ejecting matter with a constant power; thus the shells initially have the same kinetic energy. Lower panel: {\em PLUTO} model without energy dissipation in the hydrodynamic simulations. Same format as \figu{panel_ref}.}
\end{figure*}

\begin{table*}
	\centering
	\caption{Same as \tabl{model_results}, but for an engine ejecting shells with a constant luminosity ($\dot{E} = const.$) instead of a constant mass outflow (labeled Reference E) and a modified version of {\em PLUTO} model in which the PLUTO simulations don't account for energy dissipation. In both cases, all of the internal energy is radiated.}
	\label{tab:equal_mass_energy}
	\begin{tabular}{ l | c c c c |c c c }
		\hline
				&$\epsilon$ [\%]  	& $t_{\nu}$ [ms] 	& $N_{\mathrm{coll}} $ 	& $E_{\mathrm{eng}}$ [$10^{54}$ erg]& $E_{p, \mathrm{tot}}^{\mathrm{iso}} / E_{\gamma, \mathrm{tot}}^{\mathrm{iso}}$  & $E_{\nu, \mathrm{tot}}^{\mathrm{iso}} / E_{\gamma, \mathrm{tot}}^{\mathrm{iso}}$ & $E_{p,\mathrm{max}}$ [$10^{12}$ GeV] \\ \hline
		Reference  	&  35.8 $\pm$ 1.4 	&  55.2 $\pm$ 1.3 	&  970.1 $\pm$ 3.3 	&  1.75 $\pm$ 0.07 	&  0.42 $\pm$ 0.03 &  0.29 $\pm$ 0.05		&  1.2 $\pm$ 0.4  \\ 
		Reference E  	&  25.8 $\pm$ 1.0 	&  54.5 $\pm$ 0.5 	&  987.5 $\pm$ 2.8 	&  2.4 $\pm$ 0.1 	&  0.27 $\pm$ 0.01 &  0.080 $\pm$ 0.006 	&  0.26 $\pm$ 0.02  \\ 
		PLUTO no diss.	&  45.3 $\pm$ 2.9 	&  41.2 $\pm$ 1.1 	&  1270 $\pm$ 19 	&  1.38 $\pm$ 0.09 	&  0.54 $\pm$ 0.16 &  0.26 $\pm$ 0.04 		&  1.29 $\pm$ 0.30  \\ 
		\hline
	\end{tabular}
\end{table*}

\begin{figure*}
	\centering
	\includegraphics[width = 0.5  \linewidth]{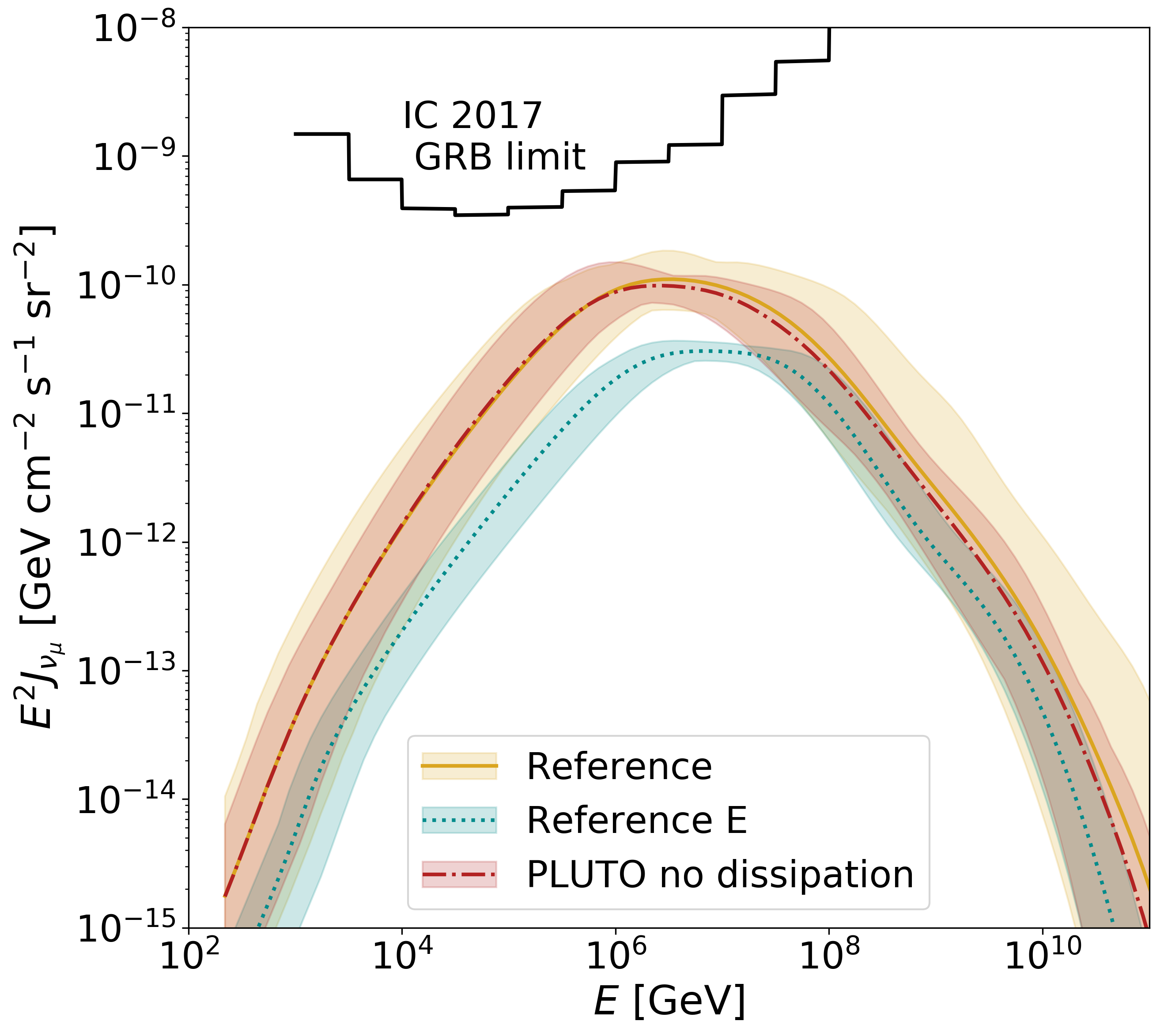}
	\caption{Comparison of the expected neutrino fluxes, assuming a rate of 667 identical GRBs per year, excluding the sub-photospheric extrapolations. Curves correspond to the average, the shaded area to the maximum / miminum neutrino fluxes achieved in the simulations. \label{fig:nufluxadd}}
\end{figure*}

\section{Additional scenarios}	
\label{app:cases}

For completeness, we discuss three additional fireball scenarios: (1) a model corresponding to GRB~1 in \citet{Bustamante:2016wpu} (the {\em Reference} model, but using a source emitting with a constant luminosity), (2) a {\em PLUTO} model, where energy dissipation is switched off in the hydrodynamic modeling and all of the internal energy is converted into radiation and (3) a model with a reduced number of initial shells, with otherwise same parameters as the {\em Reference} model.
\newpage
\subsection{Equal energy case}
\label{app:equal_energy}

The change from equal mass to equal energy, as in \citet{Bustamante:2016wpu}, has a few relevant implications. For comparison a panel for the equal energy setup is shown in \figu{panel_ref_E}.\footnote{Note that we since corrected for a factor $(1 + z)$, which is why the neutrino output in the source frame is lower by a factor of $3$. This correction does not affect the fluxes shown in the observer frame.}
The main difference is visible in the middle panel: For the equal energy assumption, the bulk of the collisions occurs at smaller radii, also a smaller spread in the dissipated energy can be observed. When considering the overall efficiency, the equal-energy scenario is significantly less efficient in converting the fireball kinetic energy into radiation. This effect can be understood by looking at the efficiency of single collisions:
If two shells of equal mass collide, their kinetic energy is converted much more efficiently into internal energy than for shells of equal energy (for high $\Gamma_r / \Gamma_s$ the single-collision efficiency approaches 1 for equal mass shells and 0.3 for equal energy shells \citep{Kino:2004uf}, the increased single-collision efficiency also is reflected in a higher overall efficiency, see Table~\ref{tab:equal_mass_energy}).   
Due to the normalization we apply, a lower overall efficiency translates into an increase of the required engine power and, subsequently, initial shell masses. More massive shells are more dense, therefore the fireball reaches the optically thin regime at larger radii and less collisions at small radii are superphotospheric. Those collisions are efficient neutrino emitters (see \citet{Bustamante:2016wpu}), which decreases the neutrino flux in the equal-energy scenario with respect to the equal-mass scenario (see also \figu{nufluxadd}).    
The applied normalization also increases the magnetic field as well as the optical depths (the first one due to the increase of $E_{\mathrm{int}}$ the latter due to the increase of the shell masses). As the optical depth to p$\gamma$-reactions limits the maximum proton energy in the collisions, higher values of $\tau_{p\gamma}$ decrease the observed $E_{p, \mathrm{max}}$, see \figu{panel_ref_E}. 
Comparing the equal energy case with the models presented in the main text, we conclude that the engine behavior (constant power vs. constant mass outflow) has a much higher impact on the observed particle fluxes than the choice of collision model. 
\subsection{PLUTO without energy dissipation}

As an other addition to the main text, we here discuss a {\em PLUTO} model where energy dissipation is not included in the hydrodynamic simulations ($\eta^* = 0 $ in \equ{ediss_pluto}). 
This corresponds to a theoretical approach where the production of internal energy is completed before particle acceleration and energy dissipation start. 
After perfoming the hydrodynamic simulation, we calculate the internal energy of the merged shell as $E_{\mathrm{int}} = E_{\mathrm{kin, before}} - E_{\mathrm{kin, after}}$, where the $E_{\mathrm{kin}}$ is the summed kinetic energy of all shells. The dissipated energy is then given by $E_{\mathrm{diss}} = E_{\mathrm{int}}$ (thus the efficiency is given by $\eta = 1.0$). In contrast to the {\em PLUTO} model in \sect{alternative_models}, the width of the emitting shell is given by \equ{shell_compression}. Again, we don't take thermal expansion after the collisions into account since the internal energy that could potentially power this effect goes into radiation. 

As far as the {\em Reference} model is concerned , we can confirm the expected neutrino fluxes for the same gamma-ray luminosity. With the given normalisation, also the evolution of the magnetic field, the optical thickness to p$\gamma$ reactions and the maximum proton energies show similar behavior. This once more strengthens our statement about the validity of the standard collision model. 
However, an increased amount of collisions ($22  \pm  1$~\% of all collisions resulted in two post-collision shells) increases the overall efficiency of the fireball by roughly $10$~\%. This is consistent with \citet{Kobayashi:2001iq} who increase the efficiency of their bursts by the almost complete thermalization of the system through multiple shell collisions. The result also once more confirms, that the probability of two post-collision shells increases, if energy dissipation is lower or not at all included in the hydrodynamic simulations.

\subsection{Reduced number of shells}
\begin{figure*}
	\centering
	\makebox[\textwidth][c]{
		\subfloat[(a)]{\includegraphics[width=.33 \textwidth]{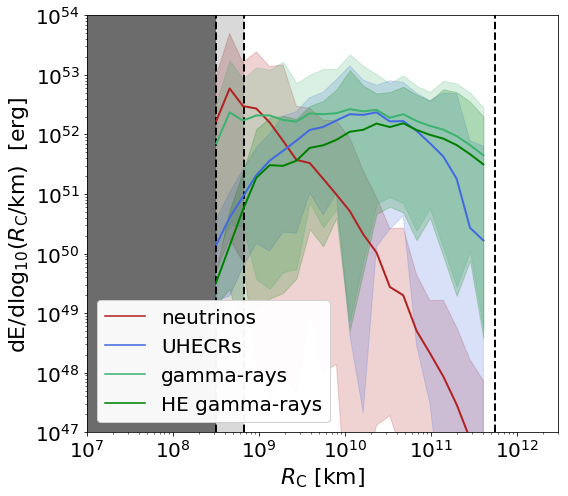}}
		\subfloat[(b)]{\includegraphics[width=.33 \textwidth]{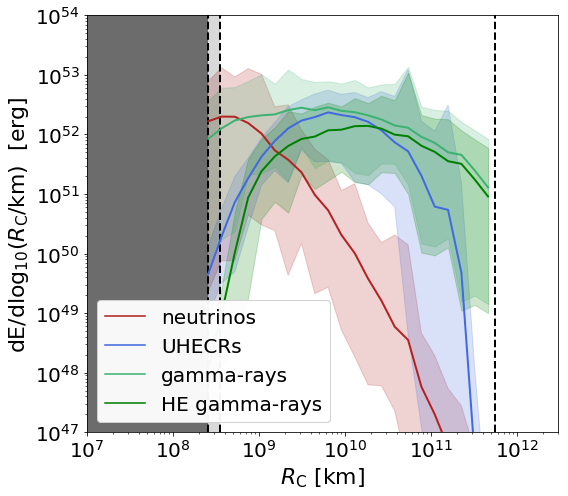}}
		\subfloat[(c)]{\includegraphics[width=.33 \textwidth]{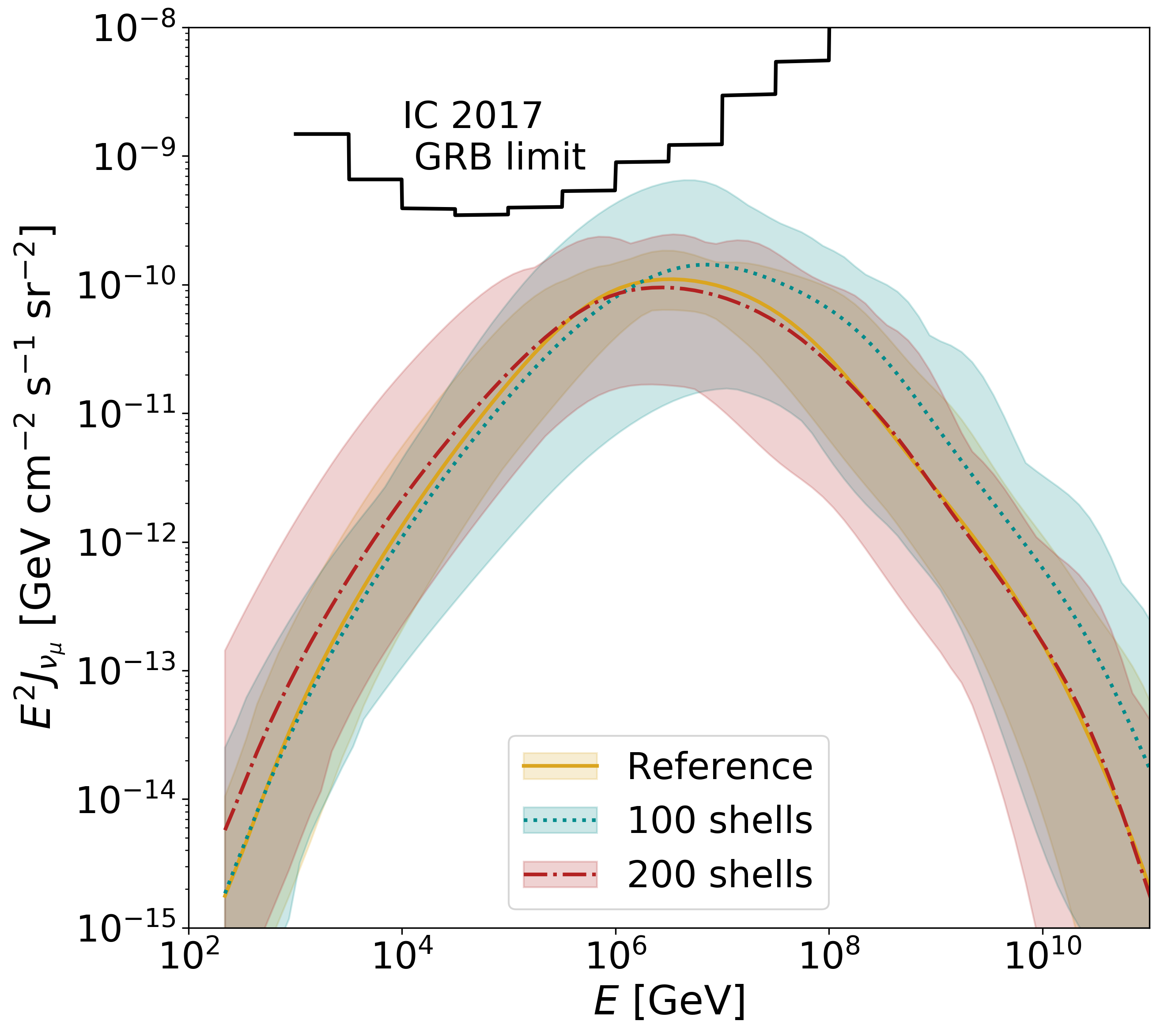}}
		
	}
	 
	\caption{\label{fig:less_shells} Energy per particle type as a function of radius and the expected neutrino fluxes for the {\em Reference} model, 
	with (a) 100 and (b) 200 initial shells. In panel (c) the neutrino fluxes are compared to our standard assumption with 1000 initial shells.}
\end{figure*}
\label{sec:reduce_number_of_shells}
To directly compare with \citep{Bustamante:2014oka,Bustamante:2016wpu} we assumed 1000 initial shells throughout the simulations discussed in the main text. However, this assumption results in variability timescales close to the fastest ever observed (see discussion in \Sec~\ref{sec:alternative_models}). We qualitatively discuss here the impact of a smaller choice of initial shells on the simulations.

Since the rate and spatial distribution of collisions directly impact the light curve it will look qualitatively different. The shell number can be reduced by either staring with a smaller jet extension or by increasing the separation between shells without changing the total length. The reduction of the initial jet size will result in shorter burst durations, typically dominated by only a few bright pulses. Longer quiescent phases between single peaks are expected for larger shell widths and separations. \figu{less_shells} shows the emission of messengers for 100 and 200 initial shells, respectively. Both configurations demonstrate that the production radii of different particles as well as the predicted neutrino fluxes (when normalized to the same gamma-ray luminosity in the optically thin regime) are similar to the {\em Reference} model with 1000 initial shells. The large statistical fluctuations for 100 initial shells result in higher sample variabilities of the predicted particle fluxes. Despite wider bands of the neutrino fluxes, and hence larger burst-to-burst variations, the model does not exceed the limits by IceCube and the conclusions derived with the choice of 1000 initial shells.
\end{appendices}
\end{document}